\documentclass[12pt,letterpaper]{article}
\usepackage{epsfig,rotating,setspace,latexsym,amsmath,epsf,amssymb,bm,theorem}
\usepackage{cite,appendix}

\usepackage{amsfonts}

\title{Multi-receiver Wiretap Channel with Public and Confidential Messages\thanks{This work
was supported by NSF Grants CCF 07-29127, CNS 09-64632, CCF
09-64645 and CCF 10-18185, and presented in part at the Allerton
Conference on Communications, Control and Computing, Monticello,
IL, September 2010.}}

\author{Ersen Ekrem \qquad Sennur Ulukus \\
\normalsize Department of Electrical and Computer Engineering\\
\normalsize University of Maryland, College Park, MD 20742 \\
\normalsize {\it ersen@umd.edu} \qquad {\it ulukus@umd.edu}}

\newcommand{\dtildew}{\tilde{\tilde{w}}}

\newcommand{\dtildeR}{\tilde{\tilde{R}}}

\newcommand{\brho}{\bm \rho}

\newcommand{\bbsigma}{\bm \Sigma}

\newcommand{\bbi}{{\mathbf{I}}}
\newcommand{\bzero}{{\mathbf{0}}}
\newcommand{\bbv}{{\mathbf{V}}}
\newcommand{\bv}{{\mathbf{v}}}

\newcommand{\bbh}{{\mathbf{H}}}

\newcommand{\bbk}{{\mathbf{K}}}
\newcommand{\bbz}{{\mathbf{Z}}}
\newcommand{\bbn}{{\mathbf{N}}}

\newcommand{\bba}{{\mathbf{A}}}
\newcommand{\bbd}{{\mathbf{D}}}

\newcommand{\bbt}{{\mathbf{T}}}
\newcommand{\bbb}{{\mathbf{B}}}

\newcommand{\bbs}{{\mathbf{S}}}
\newcommand{\bu}{{\mathbf{u}}}

\newcommand{\bbj}{{\mathbf{J}}}
\newcommand{\bbu}{{\mathbf{U}}}
\newcommand{\bx}{{\mathbf{x}}}
\newcommand{\bbx}{{\mathbf{X}}}

\newcommand{\bby}{{\mathbf{Y}}}

\newtheorem{Theo}{Theorem}

\newtheorem{Lem}{Lemma}
\newtheorem{Cor}{Corollary}
\newtheorem{Def}{Definition}

   \theoremstyle{example}

\begin{document}


\maketitle

\begin{abstract}
We study the multi-receiver wiretap channel with public and
confidential messages. In this channel, there is a transmitter
that wishes to communicate with two legitimate users in the
presence of an external eavesdropper. The transmitter sends a pair
of public and confidential messages to each legitimate user. While
there are no secrecy constraints on the public messages,
confidential messages need to be transmitted in perfect secrecy.
We study the discrete memoryless multi-receiver wiretap channel as
well as its Gaussian multi-input multi-output (MIMO) instance.
First, we consider the degraded discrete memoryless channel, and
obtain an inner bound for the capacity region by using an
achievable scheme that uses superposition coding and binning.
Next, we obtain an outer bound, and show that this outer bound
partially matches the inner bound, providing a partial
characterization for the capacity region of the degraded channel
model. Second, we obtain an inner bound for the general, not
necessarily degraded, discrete memoryless channel by using
Marton's inner bound, superposition coding, rate-splitting and
binning. Third, we consider the degraded Gaussian MIMO channel,
and show that, to evaluate both the inner and outer bounds,
considering only jointly Gaussian auxiliary random variables and
channel input is sufficient. Since the inner and outer bounds
partially match, these sufficiency results provide a partial
characterization of the capacity region of the degraded Gaussian
MIMO channel. Finally, we provide an inner bound for the capacity
region of the general,  not necessarily degraded, Gaussian MIMO
multi-receiver wiretap channel.

\end{abstract}

\setstretch{1.2}

\newpage

\section{Introduction}

Information theoretic secrecy is initiated by Wyner
in~\cite{Wyner}, where he introduces the wiretap channel and
obtains the capacity-equivocation region of a special class of
wiretap channels. Wyner considers the degraded wiretap channel,
where the eavesdropper's observation is a degraded version of the
legitimate user's observation. His result is generalized to
arbitrary, {\it not necessarily degraded}, wiretap channels
in~\cite{Korner}. Recently, the wiretap channel gathered a renewed
interest, and many multi-user extensions of the wiretap channel
have been considered. One particular multi-user extension of the
wiretap channel, that is relevant to our work here, is the
multi-receiver wiretap channel considered
in~\cite{Khandani,Ekrem_Ulukus_Asilomar08,Ekrem_Ulukus_BC_Secrecy}.
A recent survey on the secure broadcasting problem (including the
multi-receiver wiretap channel) can be found
in~\cite{secure_broad_survey}.

\begin{figure*}[t]
\vspace{-1.5cm}
\begin{center}
\includegraphics[width=5.0in]{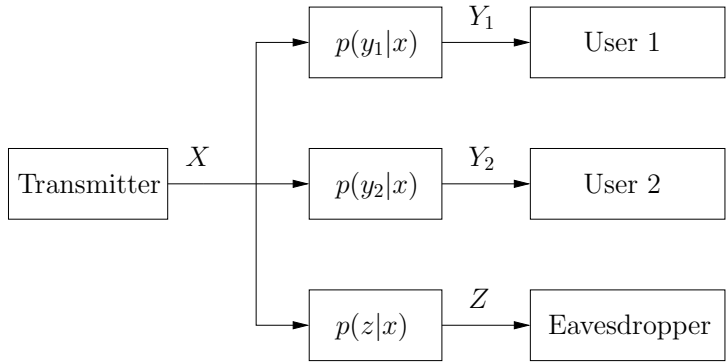}
\caption{Multi-receiver wiretap channel.} \label{mrwt_generic}
\end{center}
\end{figure*}

\begin{figure*}[t]
\vspace{-2.5cm}
\begin{center}
\includegraphics[width=5.75in]{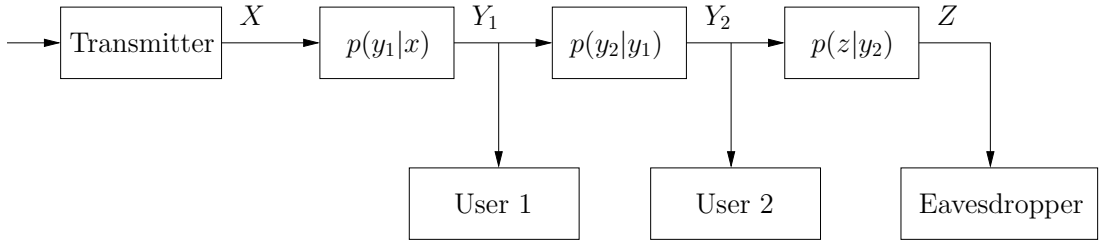}
\caption{Degraded multi-receiver wiretap channel.} \label{generic}
\end{center}
\end{figure*}

In the multi-receiver wiretap channel (see
Figure~\ref{mrwt_generic}) different from the basic wiretap
channel in~\cite{Wyner,Korner}, there are multiple legitimate
users to which the transmitter sends confidential messages in the
presence of an external eavesdropper. These multiple confidential
messages need to be kept secret from the eavesdropper. References
\cite{Khandani,Ekrem_Ulukus_Asilomar08,Ekrem_Ulukus_BC_Secrecy}
consider the degraded multi-receiver wiretap channel (see
Figure~\ref{generic}), where the observations of the legitimate
users and the eavesdropper are arranged according to a
degradedness order.
References~\cite{Khandani,Ekrem_Ulukus_Asilomar08,Ekrem_Ulukus_BC_Secrecy}
study the scenario where the transmitter sends a confidential
message to each legitimate user where these confidential messages
need to be kept perfectly secret from the eavesdropper. The
capacity region of the degraded multi-receiver wiretap channel for
this scenario is obtained in~\cite{Khandani} for two legitimate
users, and
in~\cite{Ekrem_Ulukus_Asilomar08,Ekrem_Ulukus_BC_Secrecy} for an
arbitrary number of legitimate users.

We note that to ensure the confidentiality of the messages in a
multi-receiver wiretap channel, each confidential message needs to
be randomized by many dummy
messages~\cite{Ekrem_Ulukus_Asilomar08,Ekrem_Ulukus_BC_Secrecy}.
These dummy messages protect the confidential messages from the
eavesdropper, and are decoded by the legitimate users in addition
to the confidential messages they receive. Hence, indeed, the
actual transmission rates are greater than the confidential
message rates. This also implies that the use of dummy messages
can be viewed as a {\it waste} of resources since some of the
achievable rate is spent on transmitting them. To overcome this
waste of resources, these dummy messages can be replaced with some
{\it public messages} on which there are no secrecy constraints,
as we do in this paper.

In this paper, both to overcome this waste of resources and also
to understand the actual possible transmission rates in a
multi-receiver wiretap channel, we consider the scenario where the
transmitter sends a pair of public and confidential messages to
each legitimate user. While there are no secrecy concerns on the
public messages, confidential messages need to be transmitted in
perfect secrecy. We call the channel model arising from this
scenario the {\it multi-receiver wiretap channel with public and
confidential messages.}

The capacity region of the multi-receiver wiretap channel with
public and confidential messages is closely related to the
capacity-equivocation region of the multi-receiver wiretap
channel. In the multi-receiver wiretap channel studied
in~\cite{Khandani,Ekrem_Ulukus_Asilomar08,Ekrem_Ulukus_BC_Secrecy}
(see also~\cite{MIMO_BC_Secrecy} for its Gaussian MIMO instance)
each legitimate user receives only a single message which needs to
be kept perfectly secret from the eavesdropper. That is,
\cite{Khandani,Ekrem_Ulukus_Asilomar08,Ekrem_Ulukus_BC_Secrecy,MIMO_BC_Secrecy}
consider perfect secrecy for the messages. As in the single-user
wiretap channel~\cite{Wyner,Korner}, one can be interested in the
rates \emph{and} equivocations of both messages simultaneously,
i.e., in the simultaneously achievable quadruples
$(R_1,R_{1e},R_2,R_{2e})$. This would be a four dimensional
region, as in the model discussed in this paper, where we have the
rates $R_{p1},R_{s1},R_{p2},R_{s2}$ for the rates of the
confidential and public messages of the two users. Unfortunately,
unlike the single-user wiretap channel~\cite{Wyner,Korner}, where
$(R_p,R_s)$ and $(R,R_e)$ have a one-to-one
relationship~\cite{Ruoheng_equivocation,MIMO_Equivocation,csiszar_book,biao_chen_equi,MIMO_Equivocation_Conf},
we do not have a one-to-one relationship between
$(R_{p1},R_{s1},R_{p2},R_{s2})$ and $(R_1,R_{1e},R_2,R_{2e})$. In
this paper, we focus on the public and confidential messages
framework, and focus on the region
$(R_{p1},R_{s1},R_{p2},R_{s2})$. Please see
Section~\ref{sec:equivocation} for a more detailed treatment of
the relationship between this region and the capacity-equivocation
region of the multi-receiver wiretap channel that consists of the
quadruples $(R_1,R_{1e},R_2,R_{2e})$.

In this paper, we first consider the degraded discrete memoryless
multi-receiver wiretap channel. We propose an inner bound for the
capacity region of the discrete memoryless channel. This inner
bound is based on an achievable scheme that combines superposition
coding~\cite{cover_book} and binning. Binning has been used
previously for the single-receiver and multi-receiver wiretap
channels
in~\cite{Wyner,Korner,Khandani,Ekrem_Ulukus_Asilomar08,Ekrem_Ulukus_BC_Secrecy}
to associate each confidential message with many codewords, and
hence to provide randomness for the confidential message to
protect it from the eavesdropper. In other words, by means of
binning, the confidential message is embedded into a doubly
indexed codeword where one index denotes the confidential message
and the other index (dummy index) denotes the necessary randomness
to ensure the confidentiality of the message. This second index
(dummy index) does not carry any information content.

Since in our channel model there are public messages, on which
there are no secrecy constraints, the protection of the
confidential messages from the eavesdropper can be accomplished by
using these public messages instead of the dummy messages. Thus,
the difference of binning used here from the binning used
in~\cite{Khandani,Ekrem_Ulukus_Asilomar08,Ekrem_Ulukus_BC_Secrecy}
is that, here, the confusion messages carry information, although
there are no security guarantees on this information.
Consequently, the injection of public messages into the
multi-receiver wiretap channel can be viewed as an effort to use
the {\it wasted} transmission rate due to no-information bearing
dummy indices, since these dummy indices are now replaced with
information-bearing public messages on which there are no secrecy
guarantees.

Next, we propose an outer bound for the capacity region of the
degraded discrete memoryless channel. We obtain this outer bound
by combining the converse proof techniques for the broadcast
channel with degraded message sets~\cite{BC_Degraded_Message_Set}
and the broadcast channel with confidential
messages~\cite{Korner}. This outer bound partially matches the
inner bound we propose, and therefore, it provides a partial
characterization of the capacity region of the degraded discrete
memoryless channel. In particular, when we specialize these inner
and outer bounds by setting either the public message rate of the
second legitimate user or the confidential message rate of the
first legitimate user to zero, they match and provide the exact
capacity region for these two scenarios. Moreover, when we set the
rates of both of the public messages to zero, these inner and
outer bounds match, and yield the secrecy capacity region of the
degraded discrete memoryless channel.

Second, we consider the general, not necessarily degraded,
discrete memoryless multi-receiver wiretap channel. We propose an
inner bound for the capacity region of the general channel by
using Marton's inner bound~\cite{Marton}, superposition coding,
rate-splitting and binning. This inner bound generalizes the inner
bound we proposed for the degraded case by using Marton's coding.

Third, we consider the degraded Gaussian multi-input multi-output
(MIMO) instance of this channel model. This generalizes our work
in~\cite{MIMO_BC_Secrecy}, where we consider the general, not
necessarily degraded, Gaussian MIMO channel only with confidential
messages. For the degraded Gaussian MIMO channel in this paper, we
first show that it is sufficient to consider jointly Gaussian
auxiliary random variables and channel input for the evaluation of
the inner bound we proposed for the degraded discrete memoryless
channel. In other words, we prove that there is no other possible
selection of auxiliary random variables and channel input which
can provide a rate vector outside the Gaussian rate region that is
obtained by using jointly Gaussian auxiliary random variables and
channel input. We prove the sufficiency of Gaussian auxiliary
random variables and channel input by using the de Bruijn
identity~\cite{Blachman,Palomar_Gradient}, a differential
relationship between the differential entropy and the Fisher
information matrix, in conjunction with the properties of the
Fisher information matrix.

Next, we consider the outer bound we proposed for the degraded
discrete memoryless channel. We show that, similar to the inner
bound, considering only jointly Gaussian auxiliary random
variables and channel input is sufficient to evaluate this outer
bound for the degraded Gaussian MIMO channel. Indeed, this
sufficiency result is already implied by the sufficiency of
jointly Gaussian auxiliary random variables and channel input for
the inner bound, because of the partial match between the inner
and the outer bounds. Moreover, this partial match also gives us a
partial characterization of the capacity region of the degraded
Gaussian MIMO channel. The inner and outer bounds for the degraded
Gaussian MIMO channel completely match giving us the exact
capacity region, when either the public message rate of the second
legitimate user or the confidential message rate of the first
legitimate user is zero. Moreover, these inner and outer bounds
match for the secrecy capacity region of the degraded Gaussian
MIMO channel, which we obtain when we the rates of both of the
public messages are zero~\cite{Vector_Costa_EPI,MIMO_BC_Secrecy}.

Finally, we consider the general, not necessarily degraded,
Gaussian MIMO multi-receiver wiretap channel. We propose an inner
bound for the capacity region of the general Gaussian MIMO
channel. We obtain this inner bound by using the achievable scheme
we proposed for the general discrete memoryless channel. In
particular, we evaluate this achievable scheme by using
dirty-paper coding~\cite{Wei_Yu} to obtain an inner bound for the
capacity region of the general Gaussian MIMO channel.

\section{Discrete Memoryless Multi-Receiver Wiretap Channels}
\label{sec:discrete_model}

In this section, we study discrete memoryless multi-receiver
wiretap channels which consist of a transmitter with input
alphabet $\mathcal{X}$, two legitimate users with output alphabets
$\mathcal{Y}_1,\mathcal{Y}_2$, and an eavesdropper with output
alphabet $\mathcal{Z}$. The channel is memoryless with a
transition probability $p(y_1,y_2,z|x)$, where $X\in\mathcal{X}$
is the channel input, and
$Y_1\in\mathcal{Y}_1,Y_2\in\mathcal{Y}_2,Z\in\mathcal{Z}$ denote
the channel outputs of the first legitimate user, the second
legitimate user, and the eavesdropper, respectively.

We consider the scenario in which, the transmitter sends a pair of
public and confidential messages to each legitimate user. While
there are no secrecy constraints on the public messages, the
confidential messages need to be transmitted in perfect secrecy.
We call the channel model arising from this scenario the {\it
multi-receiver wiretap channel with public and confidential
messages}.

An $(n,2^{nR_{p1}},2^{nR_{s1}},2^{nR_{p2}},2^{nR_{s2}})$ code for
this channel consists of four message sets,
$\mathcal{W}_{p1}=\{1,\ldots,2^{nR_{p1}}\}$,
$\mathcal{W}_{s1}=\{1,\ldots,2^{nR_{s1}}\}$,
$\mathcal{W}_{p2}=\{1,\ldots,2^{nR_{p2}}\}$,
$\mathcal{W}_{s2}=\{1,\ldots,2^{nR_{s2}}\}$, one encoder at the
transmitter $f:\mathcal{W}_{p1}\times
\mathcal{W}_{s1}\times\mathcal{W}_{p2}\times
\mathcal{W}_{s2}\rightarrow \mathcal{X}^{n}$, and one decoder at
each legitimate user $g_j:\mathcal{Y}_j^n\rightarrow
\mathcal{W}_{pj}\times\mathcal{W}_{sj}$, for $j=1,2$. The
probability of error is defined as
$P_e^n=\max\{P_{e,1}^n,P_{e,2}^n\}$, where
$P_{e,j}^n=\Pr[g_j(Y_j^n)\neq (W_{pj},W_{sj})]$, for $j=1,2,$ and
$W_{p1},W_{s1},W_{p2},W_{s2}$ are uniformly distributed random
variables in
$\mathcal{W}_{p1},\mathcal{W}_{s1},\mathcal{W}_{p2},\mathcal{W}_{s2}$,
respectively. A rate tuple $(R_{p1},R_{s1},R_{p2},R_{s2})$ is said
to be achievable if there exists an
$(n,2^{nR_{p1}},2^{nR_{s1}},2^{nR_{p2}}, 2^{nR_{s2}})$ code which
satisfies $\lim_{n\rightarrow \infty}P_{e}^n=0$ and
\begin{align}
\lim_{n\rightarrow \infty}\frac{1}{n} I(W_{s1},W_{s2};Z^n)=0
\label{perfect_secrecy}
\end{align}
We note that the perfect secrecy requirement given by
(\ref{perfect_secrecy}) implies the following individual perfect
secrecy requirements
\begin{align}
\lim_{n\rightarrow \infty}\frac{1}{n} I(W_{s1};Z^n)=0 \quad
\textrm{and} \quad \lim_{n\rightarrow \infty}\frac{1}{n}
I(W_{s2};Z^n)=0
\end{align}

The capacity region of the multi-receiver wiretap channel with
public and confidential messages is defined as the convex closure
of all achievable rate tuples $(R_{p1},R_{s1},R_{p2},R_{s2})$, and
is denoted by $\mathcal{C}$.

In Section~\ref{sec:degraded}, we study {\it degraded}
multi-receiver wiretap channels which satisfy the following Markov
chain
\begin{align}
X\rightarrow Y_1\rightarrow Y_2\rightarrow Z
\end{align}

\subsection{Capacity-Equivocation Region of the Multi-Receiver Wiretap Channel}
\label{sec:equivocation}

In this section, we consider the capacity-equivocation region of
the multi-receiver wiretap channel and discuss its connection to
the capacity region of the multi-receiver wiretap channel with
public and confidential messages. To this end, we consider both
rates and equivocations of the messages in the multi-receiver
wiretap channel. Each legitimate user receives a single message
which needs to be kept hidden as much as possible from the
eavesdropper. In particular, each message has its own rate and a
corresponding equivocation. The equivocations are quantified by
the following conditional entropies:
\begin{align}
\frac{1}{n}H(W_1|Z^n)\quad \textrm{and}\quad
\frac{1}{n}H(W_2|Z^n)\quad \textrm{and}\quad
\frac{1}{n}H(W_1,W_2|Z^n)
\end{align}
An $(n,2^{nR_{1}},2^{nR_{e1}},2^{nR_{2}},2^{nR_{e2}})$ code for
this channel consists of two message sets,
$\mathcal{W}_{1}=\{1,\ldots,2^{nR_{1}}\}$,
$\mathcal{W}_{2}=\{1,\ldots,2^{nR_{2}}\}$, one encoder at the
transmitter $f:\mathcal{W}_{1}\times \mathcal{W}_{2}\rightarrow
\mathcal{X}^{n}$, and one decoder at each legitimate user
$g_j:\mathcal{Y}_j^n\rightarrow \mathcal{W}_{j}$, for $j=1,2$. The
probability of error is defined as
$P_e^n=\max\{P_{e,1}^n,P_{e,2}^n\}$, where
$P_{e,j}^n=\Pr[g_j(Y_j^n)\neq W_{j}]$, for $j=1,2,$ and
$W_{1},W_{2}$ are uniformly distributed random variables in
$\mathcal{W}_{1},\mathcal{W}_{2}$, respectively. A rate tuple
$(R_{1},R_{e1},R_{2},R_{e2})$ is said to be achievable if there
exists an $(n,2^{nR_{1}},2^{nR_{e1}},2^{nR_{2}}, 2^{nR_{e2}})$
code which satisfies $\lim_{n\rightarrow \infty}P_{e}^n=0$ and
\begin{align}
R_{e1}&\leq \lim_{n\rightarrow \infty }\frac{1}{n}H(W_1|Z^n)\label{def_equi_1}\\
R_{e2}&\leq \lim_{n\rightarrow \infty }\frac{1}{n}H(W_2|Z^n)\label{def_equi_2}\\
R_{e1}+R_{e2}&\leq \lim_{n\rightarrow \infty
}\frac{1}{n}H(W_1,W_2|Z^n) \label{def_equi_3}
\end{align}
The capacity-equivocation region of the multi-receiver wiretap
channel is defined as the convex closure of all achievable rate
tuples $(R_1,R_{e1},R_2,R_{e2})$, and is denoted by
$\mathcal{C}^e$.

We now have some remarks about the definition of the equivocations
by (\ref{def_equi_1})-(\ref{def_equi_3}). First, if we consider
the perfect secrecy rates, i.e., $R_{e1}=R_1,R_{e2}=R_2$, then the
last constraint in (\ref{def_equi_3}) is sufficient, since this
last constraint implies the individual constraints in
(\ref{def_equi_1})-(\ref{def_equi_2}). Our second remark is about
the sum equivocation constraint in (\ref{def_equi_3}). If we
consider the single-legitimate user case as
in~\cite{Wyner,Korner}, we have only one of the two constraints in
(\ref{def_equi_1})-(\ref{def_equi_2}) as equivocation. A direct
generalization of this definition to the multiple legitimate user
case already yields the individual constraints in
(\ref{def_equi_1})-(\ref{def_equi_2}), however here we put one
more constraint by (\ref{def_equi_3}). The reason behind this
additional sum equivocation constraint in (\ref{def_equi_3}) is as
follows. The individual constraints in
(\ref{def_equi_1})-(\ref{def_equi_2}) consider the possibility
that the eavesdropper can try to decode $W_1$ and $W_2$
separately, however, these two individual constraints in
(\ref{def_equi_1})-(\ref{def_equi_2}) do not consider the
possibility that the eavesdropper might try to decode $W_1,W_2$
jointly. Thus, to reflect the possibility that the eavesdropper
might try to decode $W_1,W_2$ jointly on the equivocations, we
include the sum equivocation constraint in (\ref{def_equi_3}).

Next, we discuss the connection between the capacity region of the
multi-receiver wiretap channel with public and confidential
messages and the capacity-equivocation region of the
multi-receiver wiretap channel. For the single-legitimate user
case, there is a one-to-one correspondence between these two
regions as stated in the following
lemma~\cite{Ruoheng_equivocation,MIMO_Equivocation,csiszar_book,biao_chen_equi,MIMO_Equivocation_Conf}.
\begin{Lem}
\label{lemma_inclusion} $(R_{p1},R_{s1},0,0)\in\mathcal{C}$ iff
$(R_1=R_{p1}+R_{s1},R_{e1}=R_{s1},0,0)\in\mathcal{C}^e$.
\end{Lem}
The ``if'' part of this lemma follows from the definitions of the
two regions. However, the ``only if'' part of this lemma uses the
properties of the capacity achieving coding scheme for the region
$\mathcal{C}_s^e$\footnote{$\mathcal{C}_s^e$ denotes the
capacity-equivocation region for the single legitimate user
case.}. Since as of now we do not know the capacity-achieving
scheme for $\mathcal{C}^e$, we can provide only a partial
generalization of Lemma~\ref{lemma_inclusion} to the multiple
legitimate user case as stated in the following lemma.
\begin{Lem}
\label{lemma_inclusion_alt} If
$(R_{p1},R_{s1},R_{p2},R_{s2})\in\mathcal{C}$, we have
$(R_1=R_{p1}+R_{s1},R_{e1}=R_{s1},R_2=R_{p2}+R_{s2},R_{e2}=R_{s2})\in\mathcal{C}^e$.
\end{Lem}
Lemma~\ref{lemma_inclusion_alt} can be proved by using the
definitions of $\mathcal{C}$ and $\mathcal{C}^e$. This lemma
states that all inner bounds we obtain for $\mathcal{C}$ also
serve as inner bounds for $\mathcal{C}^e$. Indeed, the outer
bounds we obtain for $\mathcal{C}$ and the capacity results we
obtain for the sub-regions of $\mathcal{C}$ (a sub-region of
$\mathcal{C}$ is a region that can be obtained from $\mathcal{C}$
by setting one or more of the four rates involved in $\mathcal{C}$
to zero) can be shown to be valid for $\mathcal{C}^e$.

\subsection{Degraded Channels}

\label{sec:degraded} In this section, we consider the degraded
multi-receiver wiretap channel. We first present an inner bound
for $\mathcal{C}$ in the following theorem.

\begin{Theo}
\label{theorem_inner_bound} An achievable rate region for the
multi-receiver wiretap channel with public and confidential
messages is given by the union of rate tuples
$(R_{p1},R_{s1},R_{p2},R_{s2})$ satisfying
\begin{align}
R_{s2} &\leq I(U;Y_2)-I(U;Z)\label{final_bounds_first}\\
R_{s1}+R_{s2}&\leq I(U;Y_2)+I(X;Y_1|U) -I(X;Z)\label{final_bounds_3}\\
R_{p2}+R_{s2} &\leq I(U;Y_2)\label{final_bounds_4}\\
R_{s1}+R_{p2}+R_{s2} &\leq I(U;Y_2)+I(X;Y_1|U) -I(X;Z|U)\label{extra_bound} \\
R_{p1}+R_{s1}+R_{p2}+R_{s2} &\leq I(U;Y_2)+I(X;Y_1|U)
\label{final_bounds_last}
\end{align}
where $(U,X)$ satisfy the following Markov chain
\begin{align}
U\rightarrow X\rightarrow Y_1 \rightarrow Y_2 \rightarrow
Z\label{ach_markov_chain}
\end{align}
\end{Theo}

The achievable rate region given by
Theorem~\ref{theorem_inner_bound} can be obtained from
Theorem~\ref{theorem_inner_bound_general}, which will be
introduced in the next section. Here, we provide an outline of the
proof of Theorem~\ref{theorem_inner_bound}. We prove
Theorem~\ref{theorem_inner_bound} in two steps. As a first step,
one can show that the rate tuples $(R_{p1},R_{s1},R_{p2},R_{s2})$
satisfying
\begin{align}
R_{p2}&\leq I(U;Z) \label{original_bound_first} \\
R_{s2} &\leq I(U;Y_2)-I(U;Z) \\
R_{p1} &\leq I(X;Z|U) \\
R_{s1} &\leq I(X;Y_1|U)-I(X;Z|U) \label{original_bound_last}
\end{align}
are achievable, where $(U,X)$ satisfy (\ref{ach_markov_chain}). We
show the achievability of rate tuples $(R_{p1},R_{s1},\break
R_{p2}, R_{s2})$ satisfying
(\ref{original_bound_first})-(\ref{original_bound_last}) by using
superposition coding. The differences of the superposition coding
used here from the original superposition coding that attains the
capacity region of the degraded broadcast
channel~\cite{cover_book} are that both public and confidential
messages of a legitimate user are transmitted by the same layer of
the codebook, and the public messages provide the necessary
randomness to protect the confidential messages from the
eavesdropper. In other words, in addition to their information
content, the public messages also serve as the confusion messages
that prevent the eavesdropper from decoding the confidential
messages.

We also note that the achievability of the region given in
(\ref{original_bound_first})-(\ref{original_bound_last}) can be
concluded by using the achievable scheme
in~\cite{Khandani,Ekrem_Ulukus_Asilomar08,Ekrem_Ulukus_BC_Secrecy},
which was designed for the degraded multi-receiver wiretap channel
only with confidential messages. This achievable scheme also uses
superposition coding and binning. In this scheme, to achieve the
confidential message rates
\begin{align}
R_{s2} &= I(U;Y_2)-I(U;Z) \\
R_{s1} &= I(X;Y_1|U)-I(X;Z|U)
\end{align}
each confidential message rate is equipped with the rate of some
dummy messages, which provide the necessary protection for the
confidential messages against the eavesdropper. In particular, the
confidential message rates $R_{s2}$ and $R_{s1}$ are equipped with
the following dummy message rates
\begin{align}
\tilde{R}_{s2}&= I(U;Z) \label{dummy_rate_1}\\
\tilde{R}_{s1} &= I(X;Z|U) \label{dummy_rate_2}
\end{align}
where $\tilde{R}_{sj}$ is the dummy message rate spent for the
$j$th legitimate user's confidential message rate $R_{sj}$. Since
in our channel model there are public messages on which there are
no secrecy constraints, these dummy message rates can be used as
public message rates yielding the achievable rate region given in
(\ref{original_bound_first})-(\ref{original_bound_last}). We note
that the presence of public messages in our channel model prevents
the waste of resources due to the transmission of the dummy
messages at rates given by
(\ref{dummy_rate_1})-(\ref{dummy_rate_2}) to protect the
confidential messages.

As the second step to prove Theorem~\ref{theorem_inner_bound}, one
can use the following facts
\begin{itemize}

\item since confidential messages can be considered as public
messages as well, each legitimate user's confidential message rate
$R_{sj}$ can be given up in favor of its public message rate
$R_{pj}$, i.e., if $(R_{p1},R_{s1},R_{p2},R_{s2})$ is achievable,
$(R_{p1}+\alpha_1,R_{s1}-\alpha_1,R_{p2}+\alpha_2,R_{s2}-\alpha_2)$
is also achievable for any non-negative $(\alpha_1,\alpha_2)$
pairs satisfying $\alpha_j \leq R_{sj}$,

\item since the channel is degraded, the second legitimate user's
confidential message rate $R_{s2}$ can be given up in favor of the
first legitimate user's public and confidential message rates
$R_{p1}$ and $R_{s1}$, i.e., if $(R_{p1},R_{s1},R_{p2},R_{s2})$ is
achievable,
$(R_{p1}+\alpha,R_{s1}+\beta,R_{p2},R_{s2}-\alpha-\beta)$ is also
achievable for any non-negative $(\alpha,\beta)$ pairs satisfying
$\alpha+\beta \leq R_{s2} $,

\item since the channel is degraded, the second legitimate user's
public message rate $R_{p2}$ can be given up in favor of the first
legitimate user's public message rate $R_{p1}$,  i.e., if
$(R_{p1},R_{s1},R_{p2},R_{s2})$ is achievable,
$(R_{p1}+\alpha,R_{s1},R_{p2}-\alpha,R_{s2})$ is also achievable
for any non-negative $\alpha$ satisfying $\alpha \leq R_{p2} $,
\end{itemize}
in conjunction with Fourier-Motzkin elimination, and show that the
region given in
(\ref{original_bound_first})-(\ref{original_bound_last}) is
equivalent to the one given in
(\ref{final_bounds_first})-(\ref{final_bounds_last}).

The reason we state the achievable rate region by using the bounds
in (\ref{final_bounds_first})-(\ref{final_bounds_last}) instead of
the bounds in
(\ref{original_bound_first})-(\ref{original_bound_last}) is that
the former expressions simplify the comparison of the inner bound
with the outer bound which will be introduced in the sequel.
Another reason that we state achievable region by using the bounds
in (\ref{final_bounds_first})-(\ref{final_bounds_last}) is that
they are more convenient for an explicit evaluation for the case
of degraded Gaussian MIMO channel, which will be considered in
Section~\ref{sec:degraded_Gaussian}.

Now, we introduce the following outer bound for the capacity
region of the degraded discrete memoryless multi-receiver wiretap
channel with public and confidential messages.
\begin{Theo}
\label{theorem_outer_bound} The capacity region of the degraded
multi-receiver wiretap channel with public and confidential
messages is contained in the union of rate tuples
$(R_{p1},R_{s1},R_{p2},R_{s2})$ satisfying
\begin{align}
R_{s2} &\leq I(U;Y_2)-I(U;Z)\label{outer_bounds_first}\\
R_{s1}+R_{s2}&\leq I(U;Y_2)+I(X;Y_1|U) -I(X;Z)\\
R_{p2}+R_{s2} &\leq I(U;Y_2)\\
R_{p1}+R_{s1}+R_{p2}+R_{s2} &\leq I(U;Y_2)+I(X;Y_1|U)
\label{outer_bounds_last}
\end{align}
for some $(U,X)$ such that $U,X$ exhibit the following Markov
chain
\begin{align}
U\rightarrow X \rightarrow Y_1 \rightarrow Y_2 \rightarrow Z
\end{align}
\end{Theo}

The proof of Theorem~\ref{theorem_outer_bound} is given in
Appendix~\ref{proof_of_theorem_outer_bound}. This outer bound
provides a partial converse for the capacity region of the
degraded multi-receiver wiretap channel because the only
difference between the inner bound in
Theorem~\ref{theorem_inner_bound} and the outer bound in
Theorem~\ref{theorem_outer_bound} is the bound on
$R_{s1}+R_{p2}+R_{s2}$ given by (\ref{extra_bound}). In
particular, in addition to the bounds defining the outer bound for
the capacity region, the inner bound includes the following
constraint
\begin{align}
R_{s1}+R_{p2}+R_{s2} &\leq I(U;Y_2)+I(X;Y_1|U)-I(X;Z|U)
\end{align}
Besides that, the inner and outer bounds are identical.

Despite this difference, there are cases for which the exact
capacity region can be obtained. First, we note that the inner
bound in Theorem~\ref{theorem_inner_bound} and the outer bound in
Theorem~\ref{theorem_outer_bound} match when the confidential
message rate of the first legitimate user is zero, i.e.,
$R_{s1}=0$.
\begin{Cor}
\label{cor_degraded_mrwt_case_two} The capacity region of the
degraded multi-receiver wiretap channel without the first
legitimate user's confidential message is  given by the union of
rate triples $(R_{p1},R_{p2},R_{s2})$ satisfying
\begin{align}
R_{s2}&\leq I(U;Y_2)-I(U;Z)\\
R_{s2}+R_{p2}&\leq I(U;Y_2)\\
R_{p1}+R_{p2}+R_{s2}&\leq I(U;Y_2)+I(X;Y_1|U)
\end{align}
where $U,X$ exhibit the following Markov chain
\begin{align}
U \rightarrow X \rightarrow Y_1\rightarrow Y_2  \rightarrow Z
\end{align}
\end{Cor}
Corollary~\ref{cor_degraded_mrwt_case_two} can be proved by
setting $R_{s1}=0$ in both Theorem~\ref{theorem_inner_bound} and
Theorem~\ref{theorem_outer_bound} and eliminating the redundant
bounds.

Next, we note that the inner bound in
Theorem~\ref{theorem_inner_bound} and the outer bound in
Theorem~\ref{theorem_outer_bound} match when the public message
rate of the second legitimate user is zero, i.e., $R_{p2}=0$.
\begin{Cor}
\label{cor_degraded_mrwt_case_one} The capacity region of the
degraded multi-receiver wiretap channel without the second
legitimate user's public message is given by the union of rate
triples $(R_{p1},R_{s1},R_{s2})$ satisfying
\begin{align}
R_{s2}&\leq I(U;Y_2)-I(U;Z)\\
R_{s1}+R_{s2}&\leq I(U;Y_2)+I(X;Y_1|U)-I(X;Z)\\
R_{p1}+R_{s1}+R_{s2}&\leq I(U;Y_2)+I(X;Y_1|U)
\end{align}
where $U,X$ exhibit the following Markov chain
\begin{align}
U \rightarrow X \rightarrow Y_1\rightarrow Y_2  \rightarrow Z
\end{align}
\end{Cor}
Corollary~\ref{cor_degraded_mrwt_case_one} can be proved by
setting $R_{p2}=0$ in both Theorem~\ref{theorem_inner_bound} and
Theorem~\ref{theorem_outer_bound} and eliminating the redundant
bounds.

Corollary~\ref{cor_degraded_mrwt_case_one} also implies that the
inner bound in Theorem~\ref{theorem_inner_bound} and the outer
bound in Theorem~\ref{theorem_outer_bound} match on the secrecy
capacity region of the degraded multi-receiver wiretap channel. In
particular, the inner bound in Theorem~\ref{theorem_inner_bound}
and the outer bound in Theorem~\ref{theorem_outer_bound} match if
the rates of both public messages are set to zero, i.e.,
$R_{p1}=R_{p2}=0$. The secrecy capacity region of the degraded
multi-receiver wiretap channel is given by the following
corollary.
\begin{Cor}[\!\!\cite{Khandani,Ekrem_Ulukus_BC_Secrecy,Ekrem_Ulukus_Asilomar08}]
\label{cor_sec_reg_deg_mrwt} The secrecy capacity region of the
degraded multi-receiver wiretap channel is given by the union of
rate pairs $(R_{s1},R_{s2})$ satisfying\footnote{The secrecy
capacity region of the degraded multi-receiver wiretap channel for
an arbitrary number of legitimate users can be found
in~\cite{Ekrem_Ulukus_BC_Secrecy,Ekrem_Ulukus_Asilomar08}.}
\begin{align}
R_{s2}&\leq I(U;Y_2)-I(U;Z)\\
R_{s1}+R_{s2}&\leq I(U;Y_2)+I(X;Y_1|U)-I(X;Z)
\end{align}
where $U,X$ exhibit the following Markov chain
\begin{align}
U \rightarrow X \rightarrow Y_1 \rightarrow Y_2 \rightarrow Z
\label{dummy_mc}
\end{align}
\end{Cor}

We note that in addition to its representation in
Corollary~\ref{cor_sec_reg_deg_mrwt}, the secrecy capacity region
of the degraded multi-receiver wiretap channel can be stated in an
alternative form as the union of rate pairs $(R_{s1},R_{s2})$
satisfying
\begin{align}
R_{s2}&\leq I(U;Y_2)-I(U;Z)\\
R_{s1}&\leq I(X;Y_1|U)-I(X;Z|U)
\end{align}
where $U,X$ exhibit the Markov chain in (\ref{dummy_mc}).

\subsection{General Channels}

\label{sec:inner_general} We now consider the general,  not
necessarily degraded, discrete memoryless multi-receiver wiretap
channel with public and confidential messages. We propose an inner
bound for the capacity region of the general discrete memoryless
multi-receiver wiretap channel as follows.
\begin{Theo}
\label{theorem_inner_bound_general} An achievable rate region for
the discrete memoryless multi-receiver wiretap channel with public
and confidential messages is given by the union of rate tuples
$(R_{p1},R_{s1},R_{p2},\break R_{s2})$ satisfying
\begin{align}
R_{s1} &\leq
\min_{j=1,2}I(U;Y_j|Q)+I(V_1;Y_1|U)-I(U,V_1;Z|Q) \label{ach_gen_v1_first} \\
R_{s2}&\leq \min_{j=1,2}I(U;Y_j|Q)+I(V_2;Y_2|U)-I(U,V_2;Z|Q)\\
R_{s1}+R_{s2}&\leq
\min_{j=1,2}I(U;Y_j|Q)+I(V_1;Y_1|U)+I(V_2;Y_2|U)-I(V_1;V_2|U)\nonumber\\
&\quad\quad \quad -I(U,V_1,V_2;Z|Q)\\
R_{s1}+R_{p1}&\leq \min_{j=1,2}I(U;Y_j)+I(V_1;Y_1|U)\\
R_{s2}+R_{p2}&\leq \min_{j=1,2}I(U;Y_j)+I(V_2;Y_2|U)\\
R_{s1}+R_{p1}+R_{s2}&\leq
\min_{j=1,2}I(U;Y_j)+I(V_1;Y_1|U)+I(V_2;Y_2|U)-I(V_2;Z|U)
\end{align}
\begin{align}
R_{s1}+R_{p1}+R_{s2}&\leq \min_{j=1,2}
I(U;Y_j)+2I(V_1;Y_1|U)+I(V_2;Y_2|U)-I(V_1;V_2|U) \nonumber\\
&\quad \quad \quad -I(V_1,V_2;Z|U)\\
R_{s1}+R_{s2}+R_{p2}&\leq
\min_{j=1,2}I(U;Y_j)+I(V_1;Y_1|U)+I(V_2;Y_2|U)-I(V_1;Z|U)\\
R_{s1}+R_{s2}+R_{p2}&\leq \min_{j=1,2}
I(U;Y_j)+I(V_1;Y_1|U)+2I(V_2;Y_2|U)-I(V_1;V_2|U) \nonumber\\
&\quad \quad \quad -I(V_1,V_2;Z|U)\\
R_{s1}+R_{p1}+R_{s2}+R_{p2}&\leq
\min_{j=1,2}I(U;Y_j)+I(V_1;Y_1|U)+I(V_2;Y_2|U)-I(V_1;V_2|U)
 \label{ach_gen_v1_last}
\end{align}
for some $Q,U,V_1,V_2$ such that
$p(q,u,v_1,v_2,x,y_1,y_2,z)=p(q,u)p(v_1,v_2,x|u)p(y_1,y_2,z|x)$.
\end{Theo}

The proof of Theorem~\ref{theorem_inner_bound_general} is given in
Appendix~\ref{proof_of_theorem_inner_bound_general}. We note that
if one sets $Q=\phi,V_2=U,V_1=X$ in
Theorem~\ref{theorem_inner_bound_general}, the achievable rate
region in Theorem~\ref{theorem_inner_bound_general} reduces to the
one provided in Theorem~\ref{theorem_inner_bound}. Thus, the
achievable scheme in Theorem~\ref{theorem_inner_bound_general} can
be seen as a generalization of the achievable scheme in
Theorem~\ref{theorem_inner_bound}, where we achieve this
generalization by using Marton's coding and rate-splitting in
addition to the superposition coding and binning that were already
used for the achievable scheme in
Theorem~\ref{theorem_inner_bound}.

Next, we provide an outline of the achievable scheme in
Theorem~\ref{theorem_inner_bound_general}. In this achievable
scheme, we first divide each public message $W_{pj}$ into three
parts as $W_{pj}^1,W_{pj}^2,W_{pj}^3$, where the rates of the
messages $W_{pj}^1,W_{pj}^2,W_{pj}^3$ are given by
$R_{pj}^1,R_{pj}^2,R_{pj}^3$, respectively, and
$R_{pj}=R_{pj}^1+R_{pj}^2+R_{pj}^3$. Similarly, we divide each
confidential message $W_{sj}$ into two parts as
$W_{sj}^1,W_{sj}^2$, where the rates of the messages
$W_{sj}^1,W_{sj}^2$ are given by $R_{sj}^1,R_{sj}^2$,
respectively, and $R_{sj}=R_{sj}^1+R_{sj}^2$. The first parts of
the public messages, i.e., $W_{p1}^1$ and $W_{p2}^1$, are sent
through the sequences generated by $Q$. The second parts of the
public messages, i.e., $W_{p1}^2$ and $W_{p2}^2$, and the first
parts of the confidential messages, i.e., $W_{s1}^1$ and
$W_{s2}^1$, are sent through the sequences generated by $U$. Both
legitimate receivers decode these sequences, and hence, each
legitimate receiver decodes the parts of the other legitimate
user's public and confidential messages. The last parts of each
public message and each confidential message, i.e., $W_{pj}^3$ and
$W_{sj}^2$, are encoded by the sequences generated through $V_j$.
This encoding is performed by using Marton's coding~\cite{Marton}.
Each legitimate receiver, after decoding $Q^n$ and $U^n$, decodes
the sequences $V_j^n$. The details of the proof is given in
Appendix~\ref{proof_of_theorem_inner_bound_general}.

\section{Gaussian MIMO Multi-Receiver Wiretap Channels}

Here, we consider the Gaussian MIMO multi-receiver wiretap channel
which is defined by
\begin{align}
\bby_1&=\bbx+\bbn_1 \label{channel_def_1} \\
\bby_2&=\bbx+\bbn_2 \label{channel_def_2} \\
\bbz&=\bbx+\bbn_Z \label{channel_def_3}
\end{align}
where the channel input $\bbx$ is subject to a covariance
constraint
\begin{align}
E\left[\bbx \bbx^\top \right]\preceq \bbs
\label{covariance_constraint}
\end{align}
where $\bbs\succ \bzero$ and $\bbn_1,\bbn_2,\bbn_Z$ are zero-mean
Gaussian random vectors with covariance matrices
$\bbsigma_1,\bbsigma_2,\bbsigma_Z$, respectively.

In Section~\ref{sec:degraded_Gaussian}, we consider {\it degraded}
Gaussian MIMO multi-receiver wiretap channels for which the noise
covariance matrices $\bbsigma_1,\bbsigma_2,\bbsigma_Z$ satisfy the
following order
\begin{align}
\bzero   \prec \bbsigma_1 \preceq \bbsigma_2\preceq \bbsigma_Z
\label{covariance_matrices}
\end{align}
In a multi-receiver wiretap channel, since the capacity region
depends only on the conditional marginal distributions of the
transmitter-receiver links, but not on the entire joint
distribution of the channel, the correlations among
$\bbn_1,\bbn_2,\bbn_Z$ do not affect the capacity region. Thus,
without changing the corresponding capacity region, we can adjust
the correlation structure among these noise vectors to ensure that
they satisfy the Markov chain
\begin{align}
\bbx\rightarrow \bby_1\rightarrow \bby_2\rightarrow \bbz
\label{degraded_Markov_chain}
\end{align}
which is always possible because of our assumption about the
covariance matrices in~(\ref{covariance_matrices}). Thus, for any
Gaussian MIMO multi-receiver wiretap channel satisfying the order
in (\ref{covariance_matrices}), we can assume that it also
exhibits the Markov chain in (\ref{degraded_Markov_chain}) without
changing the capacity region of the original channel.



\subsection{Comments on the Channel Model and the Covariance Constraint}
 We provide some comments about the way we define the Gaussian MIMO
 multi-receiver wiretap channel. The first one is about the
 covariance constraint in (\ref{covariance_constraint}). Though it is more common to define capacity regions
 under a total power constraint, i.e., ${\rm tr}\left(E\left[\bbx \bbx^{\top}\right]\right) \leq
 P$, the covariance constraint in
 (\ref{covariance_constraint}) is more general and it
 subsumes the total power constraint as a special case~\cite{Shamai_MIMO}.
In particular, if we denote the capacity region under the
constraint in (\ref{covariance_constraint}) by $C(\bbs)$, then the
capacity region under the trace constraint, ${\rm
tr}\left(E\left[\bbx\bbx^{\top}\right]\right)\leq P$, can be
written as~\cite{Shamai_MIMO}
\begin{align}
C^{\rm trace}(P)=\bigcup_{\bbs:{\rm tr}(\bbs)\leq P} C(\bbs)
\end{align}
Similarly, the inner and outer bounds obtained for the covariance
constraint in (\ref{covariance_constraint}) can be extended to
provide the corresponding inner and outer bounds for the total
power constraint.

The second comment is about our assumption that $\bbs$ is strictly
positive definite. This assumption does not lead to any loss of
generality because for any Gaussian MIMO multi-receiver wiretap
channel with a positive semi-definite covariance constraint, i.e.,
$\bbs \succeq \bzero$ and $|\bbs|=0$, we can always construct an
equivalent channel with the constraint $E\left[\bbx
\bbx^{\top}\right]\preceq \bbs^{\prime}$ where $\bbs^{\prime}\succ
\bzero$ (see Lemma~2 of \cite{Shamai_MIMO}), which has the same
capacity region.

The last comment is about the assumption that the transmitter and
all receivers have the same number of antennas. This assumption is
implicit in the channel definition, see
(\ref{channel_def_1})-(\ref{channel_def_3}), and also in the
definition of degradedness, see (\ref{covariance_matrices}).
However, we can extend the definition of the Gaussian MIMO
multi-receiver wiretap channel to include the cases where the
number of transmit antennas and the number of receive antennas at
each receiver are not necessarily the same by introducing the
following channel model
\begin{align}
\bby_1&=\bbh_1 \bbx+\bbn_1\label{MIMO_channel_def_again_1}\\
\bby_2&=\bbh_2 \bbx+\bbn_2\label{MIMO_channel_def_again_2}\\
\bbz&=\bbh_Z \bbx+\bbn_Z\label{MIMO_channel_def_again_3}
\end{align}
where $\bbh_1,\bbh_2,\bbh_Z $ are the channel matrices of sizes
$r_1 \times t, r_2 \times t, r_Z \times t$, respectively, and
$\bbx$ is of size $t\times 1$. The channel outputs
$\bby_1,\bby_2,\bbz$ are of sizes $r_1 \times 1, r_2 \times 1, r_Z
\times 1$, respectively. The Gaussian noise vectors
$\bbn_1,\bbn_2,\bbn_Z$ are assumed to have identity covariance
matrices.

Next, we introduce the definition of the degradedness for the
channel model given in
(\ref{MIMO_channel_def_again_1})-(\ref{MIMO_channel_def_again_3})~\cite{Tie_Liu_Compound}.
The Gaussian MIMO multi-receiver wiretap channel in
(\ref{MIMO_channel_def_again_1})-(\ref{MIMO_channel_def_again_3})
is said to be degraded (according to the Markov chain $\bbx
\rightarrow \bby_1 \rightarrow \bby_2 \rightarrow \bbz$) if the
following two conditions hold:
\begin{itemize}
\item  There is a matrix $\bbd_{21}$ which satisfies
$\bbh_2=\bbd_{21}\bbh_1$  and $\bbd_{21}\bbd_{21}^\top \preceq
\bbi$,

\item  There is a matrix $\bbd_{Z2}$ which satisfies
$\bbh_Z=\bbd_{Z2}\bbh_2$  and $\bbd_{Z2}\bbd_{Z2}^\top \preceq
\bbi$.
\end{itemize}

In the rest of the paper, we consider the channel model given in
(\ref{channel_def_1})-(\ref{channel_def_3}) instead of the channel
model given in
(\ref{MIMO_channel_def_again_1})-(\ref{MIMO_channel_def_again_3}),
which is more general. However, the inner bounds, the outer
bounds, and the capacity regions we obtain for the Gaussian MIMO
multi-receiver wiretap channel defined by
(\ref{channel_def_1})-(\ref{channel_def_3}) can be extended to
provide the inner bounds, the outer bounds, the capacity regions,
respectively, for the Gaussian MIMO multi-receiver wiretap channel
defined by
(\ref{MIMO_channel_def_again_1})-(\ref{MIMO_channel_def_again_3})
using the analysis carried out in
Section~V~of~\cite{Tie_Liu_Compound} and Section 7.1 of
\cite{MIMO_BC_Secrecy}. Thus, focusing on the channel model in
(\ref{channel_def_1})-(\ref{channel_def_3}) does not result in any
loss of generality.

\subsection{Degraded Channels}
\label{sec:degraded_Gaussian}

We first provide an inner bound for the capacity region of the
degraded Gaussian MIMO multi-receiver wiretap channel with public
and confidential messages by using
Theorem~\ref{theorem_inner_bound}. The corresponding achievable
rate region is stated in the following theorem.
\begin{Theo}
\label{theorem_inner_Gauss} An achievable rate region for the
degraded Gaussian MIMO multi-receiver wiretap channel with public
and confidential messages is given by the union of rate tuples
$(R_{p1},R_{s1},\break R_{p2},R_{s2})$ satisfying
\begin{align}
 R_{s2}&\leq \frac{1}{2}\log
\frac{|\bbs+\bbsigma_2|}{|\bbk+\bbsigma_2|}
-\frac{1}{2}\log\frac{|\bbs+\bbsigma_Z|}{|\bbk+\bbsigma_Z|}\label{dummy_first}\\
R_{s1}+R_{s2}&\leq \frac{1}{2}\log
\frac{|\bbs+\bbsigma_2|}{|\bbk+\bbsigma_2|} +
\frac{1}{2}\log \frac{|\bbk+\bbsigma_1|}{|\bbsigma_1|}  -\frac{1}{2}\log\frac{ |\bbs+\bbsigma_Z|}{|\bbsigma_Z|}\\
R_{s2}+R_{p2} &\leq \frac{1}{2}\log\frac{|\bbs+\bbsigma_2|}{|\bbk+\bbsigma_2|} \\
R_{s1}+R_{s2}+R_{p2}&\leq
\frac{1}{2}\log\frac{|\bbs+\bbsigma_2|}{|\bbk+\bbsigma_2|}+
\frac{1}{2}\log\frac{ | \bbk+\bbsigma_1|}{|\bbsigma_1|} -\frac{1}{2}\log \frac{|\bbk+\bbsigma_Z|}{|\bbsigma_Z|} \label{extra_bound_gauss}\\
R_{s1}+R_{s2}+R_{p1}+R_{p2}&\leq
\frac{1}{2}\log\frac{|\bbs+\bbsigma_2|}{|\bbk+\bbsigma_2|} +
\frac{1}{2}\log\frac{ |\bbk+\bbsigma_1|}{|\bbsigma_1|}
\label{dummy_last_xx}
\end{align}
where $\bbk$ is a positive semi-definite matrix satisfying
$\bbk\preceq \bbs$.
\end{Theo}

This achievable rate region given in
Theorem~\ref{theorem_inner_Gauss} can be obtained by evaluating
the achievable rate region in Theorem~\ref{theorem_inner_bound}
for the degraded Gaussian MIMO multi-receiver wiretap channel by
using the following selection for $U,\bbx$: i) $U$ is a zero-mean
Gaussian random vector with covariance matrix $\bbs-\bbk$, ii)
$\bbx=U+U^\prime$ where $U^\prime$ is a zero-mean Gaussian random
vector with covariance matrix $\bbk$, and is independent of $U$.
We note that besides this jointly Gaussian $(U,\bbx)$ selection,
there might be other possible $(U,\bbx)$ selections which may
yield a larger region than the one obtained by using jointly
Gaussian $(U,\bbx)$. However, we show that jointly Gaussian
$(U,\bbx)$ selection is sufficient to evaluate the achievable rate
region in Theorem~\ref{theorem_inner_bound} for the degraded
Gaussian MIMO multi-receiver wiretap channel. In other words,
jointly Gaussian $(U,\bbx)$ selection exhausts the achievable rate
region in Theorem~\ref{theorem_inner_bound} for the degraded
Gaussian MIMO multi-receiver wiretap channel. This sufficiency
result is stated in the following theorem.
\begin{Theo}
\label{theorem_suff_1} For the degraded Gaussian MIMO
multi-receiver wiretap channel, the achievable rate region in
Theorem~\ref{theorem_inner_bound} is exhausted by jointly Gaussian
$(U,\bbx)$. In particular, for any non-Gaussian $(U,\bbx)$, there
exists a Gaussian $(U^G,\bbx^G)$ which yields a larger region than
the one obtained by using the non-Gaussian $(U,\bbx)$.
\end{Theo}

Next, we provide an outer bound for the capacity region of the
degraded Gaussian MIMO multi-receiver wiretap channel. This outer
bound can be obtained by evaluating the outer bound given in
Theorem~\ref{theorem_outer_bound} for the degraded Gaussian MIMO
multi-receiver wiretap channel. This evaluation is tantamount to
finding the optimal $(U,\bbx)$ which exhausts the outer bound in
Theorem~\ref{theorem_outer_bound} for the degraded Gaussian MIMO
multi-receiver wiretap channel. We show that jointly Gaussian
$(U,\bbx)$ is sufficient to exhaust the outer bound in
Theorem~\ref{theorem_outer_bound} for the degraded Gaussian MIMO
channel. The corresponding outer bound is stated in the following
theorem.
\begin{Theo}
\label{theorem_outer_Gauss} The capacity region of the degraded
Gaussian MIMO multi-receiver wiretap channel is contained in the
union of rate tuples $(R_{p1},R_{s1},R_{p2},R_{s2})$ satisfying
\begin{align}
R_{s2}&\leq \frac{1}{2}\log
\frac{|\bbs+\bbsigma_2|}{|\bbk+\bbsigma_2|}
-\frac{1}{2}\log\frac{| \bbs+\bbsigma_Z|}{|\bbk+\bbsigma_Z|}\label{dummy_first2}\\
R_{s1}+R_{s2}&\leq \frac{1}{2}\log
\frac{|\bbs+\bbsigma_2|}{|\bbk+\bbsigma_2|} +
\frac{1}{2}\log \frac{|\bbk+\bbsigma_1|}{|\bbsigma_1|}  -\frac{1}{2}\log \frac{|\bbs+\bbsigma_Z|}{|\bbsigma_Z|}\\
R_{s2}+R_{p2} &\leq \frac{1}{2}\log\frac{|\bbs+\bbsigma_2|}{|\bbk+\bbsigma_2|} \\
R_{s1}+R_{s2}+R_{p1}+R_{p2}&\leq
\frac{1}{2}\log\frac{|\bbs+\bbsigma_2|}{|\bbk+\bbsigma_2|} +
\frac{1}{2}\log \frac{|\bbk+\bbsigma_1|}{|\bbsigma_1|}
\label{dummy_last_xx2}
\end{align}
where $\bbk$ is a positive semi-definite matrix satisfying
$\bbk\preceq \bbs$.
\end{Theo}

The proofs of Theorem~\ref{theorem_suff_1} and
Theorem~\ref{theorem_outer_Gauss} are given in
Appendix~\ref{proofs_of_degraded_Gaussian_MIMO}. We prove
Theorem~\ref{theorem_suff_1} and Theorem~\ref{theorem_outer_Gauss}
by using the de Bruijn identity~\cite{Blachman,Palomar_Gradient},
a differential relationship between differential entropy and the
Fisher information matrix, in conjunction with the properties of
the Fisher information matrix. In particular, to prove
Theorem~\ref{theorem_suff_1}, we consider the region in
Theorem~\ref{theorem_inner_bound}, and show that for any
non-Gaussian $(U,\bbx)$, there exists a Gaussian $(U^G,\bbx^G)$
which yields a larger region than the one that is obtained by
evaluating the region in Theorem~\ref{theorem_inner_bound} with
the non-Gaussian $(U,\bbx)$. We note that this proof of
Theorem~\ref{theorem_suff_1} implies the proof of
Theorem~\ref{theorem_outer_Gauss}. In particular, since the region
in Theorem~\ref{theorem_inner_bound} includes all the constraints
involved in the outer bound given in
Theorem~\ref{theorem_outer_bound}, the proof of
Theorem~\ref{theorem_suff_1} reveals that for any non-Gaussian
$(U,\bbx)$, there exists a Gaussian $(U^G,\bbx^G)$ which yields a
larger region than the one that is obtained by evaluating the
region in Theorem~\ref{theorem_outer_bound} with the non-Gaussian
$(U,\bbx)$.

We note that the only difference between the inner and the outer
bounds for the degraded Gaussian MIMO multi-receiver wiretap given
in Theorem~\ref{theorem_inner_Gauss} and
Theorem~\ref{theorem_outer_Gauss}, respectively, comes from the
bound in (\ref{extra_bound_gauss}). In other words, there is one
more constraint in the inner bound given by
Theorem~\ref{theorem_inner_Gauss} than the outer bound given by
Theorem~\ref{theorem_outer_Gauss}. This additional constraint is
\begin{align}
R_{s1}+R_{s2}+R_{p2} &\leq \frac{1}{2} \log
\frac{|\bbs+\bbsigma_2|}{|\bbk+\bbsigma_2|}+ \frac{1}{2}\log
\frac{| \bbk+\bbsigma_1|}{|\bbsigma_1|} -\frac{1}{2}\log
\frac{|\bbk+\bbsigma_Z|}{|\bbsigma_Z|}
\end{align}
Besides this constraint on $R_{s1}+R_{s2}+R_{p2}$, the inner bound
in Theorem~\ref{theorem_inner_Gauss} and the outer bound in
Theorem~\ref{theorem_outer_Gauss} are the same.

We conclude this section by providing the cases where the inner
bound in Theorem~\ref{theorem_inner_Gauss} and the outer bound in
Theorem~\ref{theorem_outer_Gauss} match. We first note that the
inner bound in Theorem~\ref{theorem_inner_Gauss} and the outer
bound in Theorem~\ref{theorem_outer_Gauss} match when the
confidential message rate of the first legitimate user is zero,
i.e., $R_{s1}=0$. The corresponding capacity region is given by
the following corollary.
\begin{Cor}
\label{cor_degraded_mrwt_gauss_case_two} The capacity region of
the degraded Gaussian MIMO multi-receiver wiretap channel without
the first legitimate user's confidential message is given by the
union of rate tuples $(R_{p1},R_{p2},R_{s2})$ satisfying
\begin{align}
R_{s2}&\leq \frac{1}{2}\log
\frac{|\bbs+\bbsigma_2|}{|\bbk+\bbsigma_2|}
-\frac{1}{2}\log\frac{|\bbs+\bbsigma_Z|}{|\bbk+\bbsigma_Z|}\\
R_{s2}+R_{p2} &\leq \frac{1}{2}\log\frac{|\bbs+\bbsigma_2|}{|\bbk+\bbsigma_2|} \\
R_{s2}+R_{p1}+R_{p2}&\leq \frac{1}{2} \log
\frac{|\bbs+\bbsigma_2|}{|\bbk+\bbsigma_2|} +
\frac{1}{2}\log\frac{ | \bbk+\bbsigma_1|}{|\bbsigma_1|}
\end{align}
where $\bbk$ is a positive semi-definite matrix satisfying
$\bbk\preceq \bbs$.
\end{Cor}

We note that Corollary~\ref{cor_degraded_mrwt_gauss_case_two} is
the Gaussian MIMO version of
Corollary~\ref{cor_degraded_mrwt_case_two} which obtains the
capacity region of the degraded discrete memoryless multi-receiver
wiretap channel without the first legitimate user's confidential
message. Corollary~\ref{cor_degraded_mrwt_gauss_case_two} can be
proved by setting $R_{s1}=0$ in both
Theorem~\ref{theorem_inner_Gauss} and
Theorem~\ref{theorem_outer_Gauss} and eliminating the redundant
bounds.

We next note that the inner bound in
Theorem~\ref{theorem_inner_Gauss} and the outer bound in
Theorem~\ref{theorem_outer_Gauss} match when the public message
rate of the second legitimate user is zero, i.e., $R_{p2}=0$. The
corresponding capacity region is stated in the following
corollary.
\begin{Cor}
\label{cor_degraded_mrwt_gauss_case_one} The capacity region of
the degraded Gaussian MIMO multi-receiver wiretap channel without
the second legitimate user's public message is given by the union
of rate tuples $(R_{p1},R_{s1},R_{s2})$ satisfying
\begin{align}
 R_{s2}&\leq \frac{1}{2}\log
\frac{|\bbs+\bbsigma_2|}{|\bbk+\bbsigma_2|}
-\frac{1}{2}\log\frac{|\bbs+\bbsigma_Z|}{|\bbk+\bbsigma_Z|}\\
 R_{s1}+R_{s2}&\leq \frac{1}{2}\log
\frac{|\bbs+\bbsigma_2|}{|\bbk+\bbsigma_2|} +
\frac{1}{2}\log\frac{ |\bbk+\bbsigma_1| }{|\bbsigma_1|}
-\frac{1}{2}\log \frac{|\bbs+\bbsigma_Z|}{|\bbsigma_Z|}
\\
R_{s1}+R_{s2}+R_{p1}&\leq \frac{1}{2} \log
\frac{|\bbs+\bbsigma_2|}{|\bbk+\bbsigma_2|}+ \frac{1}{2}\log
\frac{| \bbk+\bbsigma_1|}{|\bbsigma_1|} \label{dummy_last}
\end{align}
where $\bbk$ is a positive semi-definite matrix satisfying
$\bbk\preceq \bbs$.
\end{Cor}

We note that Corollary~\ref{cor_degraded_mrwt_gauss_case_one} is
the Gaussian MIMO version of
Corollary~\ref{cor_degraded_mrwt_case_one} which obtains the
capacity region of the degraded discrete memoryless multi-receiver
wiretap channel without the second legitimate user's public
message. Corollary~\ref{cor_degraded_mrwt_gauss_case_one} can be
proved by setting $R_{p2}=0$ in both
Theorem~\ref{theorem_inner_Gauss} and
Theorem~\ref{theorem_outer_Gauss} and eliminating the redundant
bounds.

Corollary~\ref{cor_degraded_mrwt_gauss_case_one} also implies that
the inner bound in Theorem~\ref{theorem_inner_Gauss} and the outer
bound in Theorem~\ref{theorem_outer_Gauss} match for the secrecy
capacity region of the degraded Gaussian MIMO multi-receiver
wiretap channel. In particular, the inner bound
Theorem~\ref{theorem_inner_Gauss} and the outer bound in
Theorem~\ref{theorem_outer_Gauss} match if the rates of both
public messages are set to zero, i.e., $R_{p1}=R_{p2}=0$. The
secrecy capacity region of the degraded multi-receiver wiretap
channel is given by the following corollary.
\begin{Cor}[\!\!\cite{MIMO_BC_Secrecy,Vector_Costa_EPI}]
\label{cor_sec_reg_deg_MIMO_mrwt} The secrecy capacity region of
the degraded Gaussian MIMO multi-receiver wiretap channel is given
by the union of rate pairs $(R_{s1},R_{s2})$
satisfying\footnote{The secrecy capacity region of the general,
not necessarily degraded, Gaussian MIMO multi-receiver wiretap
channel for an arbitrary number of legitimate users can be found
in~\cite{MIMO_BC_Secrecy}.}
\begin{align}
R_{s2}&\leq \frac{1}{2} \log
\frac{|\bbs+\bbsigma_2|}{|\bbk+\bbsigma_2|}
-\frac{1}{2} \log \frac{|\bbs+\bbsigma_Z|}{|\bbk+\bbsigma_Z|}\\
R_{s2}+R_{s1}&\leq \frac{1}{2} \log
\frac{|\bbs+\bbsigma_2|}{|\bbk+\bbsigma_2|}+ \frac{1}{2} \log
\frac{|\bbk+\bbsigma_1|}{|\bbsigma_1|} -\frac{1}{2} \log
\frac{|\bbs+\bbsigma_Z|}{|\bbsigma_Z|}
\end{align}
where $\bbk$ is a positive semi-definite matrix satisfying $ \bbk
\preceq \bbs$.
\end{Cor}

We note that in addition to its representation in
Corollary~\ref{cor_sec_reg_deg_MIMO_mrwt}, the secrecy capacity
region of the degraded Gaussian MIMO multi-receiver wiretap
channel can be stated in an alternative form as the union of rate
pairs $(R_{s1},R_{s2})$ satisfying
\begin{align}
R_{s2}&\leq \frac{1}{2} \log
\frac{|\bbs+\bbsigma_2|}{|\bbk+\bbsigma_2|}
-\frac{1}{2} \log \frac{|\bbs+\bbsigma_Z|}{|\bbk+\bbsigma_Z|}\\
R_{s1}&\leq \frac{1}{2} \log
\frac{|\bbk+\bbsigma_1|}{|\bbsigma_1|}-\frac{1}{2} \log
\frac{|\bbk+\bbsigma_Z|}{|\bbsigma_Z|}
\end{align}
where $\bbk$ is positive semi-definite matrix satisfying $\bbk
\preceq \bbs$.

\subsection{General Channels}

Here we consider the general, i.e., {\it not necessarily
degraded}, Gaussian MIMO multi-receiver wiretap channel with
public and confidential messages, and propose an inner bound for
the capacity region of the general Gaussian MIMO multi-receiver
wiretap channel as follows.
\begin{Theo}
\label{theorem_ach_general_Gaussian_MIMO} An achievable rate
region for the general Gaussian MIMO multi-receiver wiretap
channel with public and confidential messages is given by
\begin{align}
{\rm conv}\left(\mathcal{R}_{12}(\bbk_0,\bbk_1,\bbk_2)\cup
\mathcal{R}_{21}(\bbk_0,\bbk_1,\bbk_2)\right)
\end{align}
where $\mathcal{R}_{21}(\bbk_0,\bbk_1,\bbk_2)$ is given by the
union of rate tuples $(R_{p1},R_{s1},R_{p2},R_{s2})$ satisfying
\begin{align}
R_{s1}&\leq \min_{j=1,2}\frac{1}{2} \log
\frac{|\bbk_0+\bbk_1+\bbk_2+\bbsigma_j|}{|\bbk_1+\bbk_2+\bbsigma_j|}
+\frac{1}{2}\log
\frac{|\bbk_1+\bbsigma_1|}{|\bbsigma_1|}\nonumber\\
&\qquad\quad - \frac{1}{2} \log
\frac{|\bbk_0+\bbk_1+\bbk_2+\bbsigma_Z|}{|\bbk_1+\bbk_2+\bbsigma_Z|}
-\frac{1}{2} \log
\frac{|\bbk_1+\bbsigma_Z|}{|\bbsigma_Z|}\label{ach_general_Gauss_first}
\end{align}
\begin{align}
R_{s2}&\leq \min_{j=1,2} \frac{1}{2} \log
\frac{|\bbk_0+\bbk_1+\bbk_2+\bbsigma_j|}{|\bbk_1+\bbk_2+\bbsigma_j|}+
\frac{1}{2} \log
\frac{|\bbk_1+\bbk_2+\bbsigma_2|}{|\bbk_1+\bbsigma_2|}\nonumber\\
&\qquad \quad -\frac{1}{2}\log
\frac{|\bbk_0+\bbk_1+\bbk_2+\bbsigma_Z|}{|\bbk_1+\bbsigma_Z|}\\
R_{s1}+R_{s2}&\leq \min_{j=1,2}\frac{1}{2} \log
\frac{|\bbk_0+\bbk_1+\bbk_2+\bbsigma_j|}{|\bbk_1+\bbk_2+\bbsigma_j|}
+\frac{1}{2} \log
\frac{|\bbk_1+\bbk_2+\bbsigma_2|}{|\bbk_1+\bbsigma_2|}\nonumber\\
&\qquad\quad  +\frac{1}{2}\log
\frac{|\bbk_1+\bbsigma_1|}{|\bbsigma_1|} -\frac{1}{2}\log
\frac{|\bbk_0+\bbk_1+\bbk_2+\bbsigma_Z|}{|\bbsigma_Z|}\\
R_{s1}+R_{p1}&\leq \min_{j=1,2} \frac{1}{2} \log
\frac{|\bbs+\bbsigma_j|}{|\bbk_1+\bbk_2+\bbsigma_j|}+\frac{1}{2}\log
\frac{|\bbk_1+\bbsigma_1|}{|\bbsigma_1|}\\
R_{s2}+R_{p2}&\leq \min_{j=1,2} \frac{1}{2} \log
\frac{|\bbs+\bbsigma_j|}{|\bbk_1+\bbk_2+\bbsigma_j|}+\frac{1}{2}\log
\frac{|\bbk_1+\bbk_2+\bbsigma_2|}{|\bbk_1+\bbsigma_2|}\\
R_{s1}+R_{p1}+R_{s2}&\leq \min_{j=1,2} \frac{1}{2} \log
\frac{|\bbs+\bbsigma_j|}{|\bbk_1+\bbk_2+\bbsigma_j|}+\frac{1}{2}\log
\frac{|\bbk_1+\bbsigma_1|}{|\bbsigma_1|}\nonumber\\
&\qquad \quad +\frac{1}{2}\log
\frac{|\bbk_1+\bbk_2+\bbsigma_2|}{|\bbk_1+\bbsigma_2|}
-\frac{1}{2}\log
\frac{|\bbk_1+\bbk_2+\bbsigma_Z|}{|\bbk_1+\bbsigma_Z|}\\
R_{s1}+R_{s2}+R_{p2}&\leq \min_{j=1,2} \frac{1}{2} \log
\frac{|\bbs+\bbsigma_j|}{|\bbk_1+\bbk_2+\bbsigma_j|}+\frac{1}{2}\log
\frac{|\bbk_1+\bbk_2+\bbsigma_2|}{|\bbk_1+\bbsigma_2|}\nonumber\\
&\qquad \quad +\frac{1}{2}\log
\frac{|\bbk_1+\bbsigma_1|}{|\bbsigma_1|} -\frac{1}{2}\log
\frac{|\bbk_1+\bbsigma_Z|}{|\bbsigma_Z|}\\
R_{s1}+R_{p1}+R_{s2}+R_{p2}&\leq \min_{j=1,2} \frac{1}{2}\log
\frac{|\bbs+\bbsigma_j|}{|\bbk_1+\bbk_2+\bbsigma_j|}+\frac{1}{2}\log
\frac{|\bbk_1+\bbk_2+\bbsigma_2|}{|\bbk_1+\bbsigma_2|}\nonumber\\
&\qquad \quad +\frac{1}{2}\log
\frac{|\bbk_1+\bbsigma_1|}{|\bbsigma_1|}\label{ach_general_Gauss_last}
\end{align}
for some positive semi-definite matrices $\bbk_0,\bbk_1,\bbk_2$
satisfying $\bbk_0+\bbk_1+\bbk_2\preceq \bbs$.
$\mathcal{R}_{12}(\bbk_0,\bbk_1,\bbk_2)$ can be obtained from
$\mathcal{R}_{21}(\bbk_0,\bbk_1,\bbk_2)$ by swapping the
subscripts 1 and 2.
\end{Theo}
The proof of Theorem~\ref{theorem_ach_general_Gaussian_MIMO} is
given in
Appendix~\ref{proof_of_theorem_ach_general_Gaussian_MIMO}. We
obtain the achievable rate region in
Theorem~\ref{theorem_ach_general_Gaussian_MIMO} by evaluating the
achievable rate region given in
Theorem~\ref{theorem_inner_bound_general} with jointly Gaussian
$(Q,U,V_1,V_2,\bbx)$ having a specific correlation structure. In
particular, $Q,U$ are selected in accordance with superposition
coding, and $V_1,V_2$ are encoded by using dirty-paper
encoding~\cite{Wei_Yu}.

We note that Theorem~\ref{theorem_inner_Gauss} is a special case
of Theorem~\ref{theorem_ach_general_Gaussian_MIMO}. In other
words, the achievable rate region in
Theorem~\ref{theorem_inner_Gauss} can be obtained from the
achievable rate region in
Theorem~\ref{theorem_ach_general_Gaussian_MIMO}. To this end, one
needs to consider only the region
$\mathcal{R}_{21}(\bbk_0,\bbk_1,\bbk_2)$ with
$\bbk_2=\phi,\bbk_1=\bbk,\bbs=\bbk_0+\bbk_1$. After eliminating
the redundant bounds from the corresponding region, one can
recover the achievable rate region provided in
Theorem~\ref{theorem_inner_Gauss}.

\section{Conclusions}
We study the multi-receiver wiretap channel with public and
confidential messages. We first consider the degraded discrete
memoryless channel. We provide inner and outer bounds for the
capacity region of the degraded discrete memoryless multi-receiver
wiretap channel. These inner and outer bounds partially match.
Thus, we provide a partial characterization of the capacity region
of the degraded discrete memoryless multi-receiver wiretap channel
with public and confidential messages. Second, we provide an inner
bound for the capacity region of the general, not necessarily
degraded, multi-receiver wiretap channel. We obtain the inner
bound for the general case by using an achievable scheme that
relies on superposition coding, rate-splitting, binning and
Marton's coding. Third, we consider the degraded Gaussian MIMO
multi-receiver wiretap channel. We show that, to evaluate the
inner and outer bounds, that we already proposed for the degraded
discrete memoryless channel, for the Gaussian MIMO case, it is
sufficient to consider only the jointly Gaussian auxiliary random
variables and channel input. Consequently, since these inner and
outer bounds partially match, we obtain a partial characterization
of the capacity region of the degraded Gaussian MIMO
multi-receiver wiretap channel with public and confidential
messages. Finally, we consider the general, not necessarily
degraded, Gaussian MIMO multi-receiver wiretap channel and propose
an inner bound for the capacity region of the general Gaussian
MIMO channel.

\appendices

\section{Proof of Theorem~\ref{theorem_outer_bound}}
\label{proof_of_theorem_outer_bound}

We define the following auxiliary random variables
\begin{align}
U_i&= W_{s2} W_{p2}Y_{1}^{i-1} Z_{i+1}^n,\quad i=1,\ldots,n
\end{align}
which satisfy the following Markov chains
\begin{align}
U_i\rightarrow X_i\rightarrow Y_{1i} \rightarrow Y_{2i}
\rightarrow Z_i,\quad i=1,\ldots,n \label{aux_rvs_markov_chain}
\end{align}
due to the fact that the channel is degraded and memoryless. For
any $(n,2^{nR_{p1}},2^{nR_{s1}},2^{nR_{p2}},\break 2^{nR_{s2}})$
code achieving the rate tuple $(R_{p1},R_{s1},R_{p2},R_{s2})$, due
to Fano's lemma, we have
\begin{align}
H(W_{s2},W_{p2}|Y_2^n)&\leq n\epsilon_n \\
H(W_{s1},W_{p1}|W_{s2},W_{p2},Y_1^n) &\leq n\epsilon_n
\end{align}
where $\epsilon_n\rightarrow 0$ as $n\rightarrow \infty$.
Moreover, due to the perfect secrecy requirement in
(\ref{perfect_secrecy}), we have
\begin{align}
I(W_{s1},W_{s2};Z^n) \leq n\gamma_n
\label{perfect_secrecy_implies}
\end{align}
where $\gamma_n\rightarrow 0$ as $n\rightarrow \infty$. Next, we
obtain the following
\begin{align}
H(W_{s1},W_{s2})&\leq H(W_{s1},W_{s2}|Z^n) +n\gamma_n \label{perfect_secrecy_implies_1} \\
&\leq H(W_{s1},W_{s2},W_{p1},W_{p2}|Z^n) +n\gamma_n
\label{perfect_secrecy_implies_2}
\end{align}
where (\ref{perfect_secrecy_implies_1}) comes from
(\ref{perfect_secrecy_implies}). Similarly, we obtain the
following
\begin{align}
H(W_{s2})&\leq H(W_{s2}|Z^n) +n\gamma_n \label{perfect_secrecy_implies_3} \\
&\leq H(W_{s2},W_{p2}|Z^n) +n\gamma_n
\label{perfect_secrecy_implies_4}
\end{align}
where (\ref{perfect_secrecy_implies_3}) comes from
(\ref{perfect_secrecy_implies}). Next, we introduce the following
lemmas which will be used frequently.
\begin{Lem}\bf{(\!\!\cite[Lemma~7]{Korner})}
\label{lemma_CK_1}
\begin{align}
\sum_{i=1}^n I(T_{1,i+1}^n;T_{2i}|Q,T_2^{i-1})=\sum_{i=1}^n
I(T_{2}^{i-1};T_{1i}|Q,T_{1,i+1}^{n})
\end{align}
\end{Lem}
\begin{Lem}
\label{lemma_CK_2}
\begin{align}
I(W;T_1^n|Q)-(W;T_2^n|Q)=\sum_{i=1}^n
I(W;T_{1i}|Q,T_{1}^{i-1},T_{2,i+1}^{n})-I(W;T_{2i}|Q,T_{1}^{i-1},T_{2,i+1}^{n})
\end{align}
\end{Lem}
Lemma~\ref{lemma_CK_2} can be proved by using
Lemma~\ref{lemma_CK_1}.

First, we obtain an outer bound for the second legitimate user's
confidential message rate $R_{s2}$ as follows
\begin{align}
nR_{s2}&=H(W_{s2}) \\
&\leq H(W_{s2},W_{p2}|Z^n)+n\gamma_n \label{perfect_secrecy_implies_4_implies}\\
&=H(W_{s2},W_{p2})-I(W_{s2},W_{p2};Z^n) +n\gamma_n \\
&\leq I(W_{s2},W_{p2};Y_2^n)-I(W_{s2},W_{p2};Z^n) +n(\epsilon_n+\gamma_n)\\
&=\sum_{i=1}^n I(W_{s2},W_{p2};Y_{2i}|Y_2^{i-1},Z_{i+1}^n)-I(W_{s2},W_{p2};Z_i|Y_2^{i-1},Z_{i+1}^n)+n(\epsilon_n+\gamma_n)\label{lemma_CK_2_implies}\\
&=\sum_{i=1}^n I(W_{s2},W_{p2};Y_{2i}|Y_2^{i-1},Z_{i+1}^n,Z_i)+n(\gamma_n+\epsilon_n)\label{degradedness_implies_2}\\
&\leq \sum_{i=1}^n I(W_{s2},W_{p2},Y_1^{i-1},Y_2^{i-1},Z_{i+1}^n;Y_{2i}|Z_i)+n(\gamma_n+\epsilon_n)\\
&=\sum_{i=1}^n I(W_{s2},W_{p2},Y_1^{i-1},Z_{i+1}^n;Y_{2i}|Z_i)+n(\gamma_n+\epsilon_n)\label{degradedness_implies_2x}\\
&= \sum_{i=1}^n I(U_i;Y_{2i}|Z_i)+n(\gamma_n+\epsilon_n)\\
&= \sum_{i=1}^n I(U_i;Y_{2i})-I(U_i;Z_{i})+n(\gamma_n+\epsilon_n)
\label{degradedness_implies_3}
\end{align}
where (\ref{perfect_secrecy_implies_4_implies}) comes from
(\ref{perfect_secrecy_implies_4}), (\ref{lemma_CK_2_implies}) is
due to Lemma~\ref{lemma_CK_2}. The equalities in
(\ref{degradedness_implies_2}), (\ref{degradedness_implies_2x}),
and (\ref{degradedness_implies_3}) come from the following Markov
chains
\begin{align}
W_{s2},W_{p2},Y_2^{i-1},Z_{i+1}^n \rightarrow Y_{2i} \rightarrow
&Z_i\\
W_{s2},W_{p2},Z_{i}^n,Y_{2i}\rightarrow Y_1^{i-1}\rightarrow
&Y_2^{i-1}\\
U_i\rightarrow Y_{2i} \rightarrow & Z_i
\end{align}
which follow from the fact that the channel is degraded and
memoryless.

Next, we obtain an outer bound for the confidential message sum
rate $R_{s1}+R_{s2}$ as follows
\begin{align}
\lefteqn{n(R_{s1}+R_{s2})=H(W_{s1},W_{s2}) } \\
&\leq H(W_{s1},W_{p1},W_{s2},W_{p2}|Z^n)+n\gamma_n \label{perfect_secrecy_implies_2_implies}\\
&= H(W_{s1},W_{p1},W_{s2},W_{p2})-I(W_{s1},W_{p1},W_{s2},W_{p2};Z^n)+n\gamma_n \\
&\leq I(W_{s1},W_{p1};Y_1^n|W_{s2},W_{p2})+I(W_{s2},W_{p2};Y_2^n)-I(W_{s1},W_{p1},W_{s2},W_{p2};Z^n)\nonumber\\
&\quad +n(\gamma_n+2\epsilon_n) \\
&= I(W_{s1},W_{p1};Y_1^n|W_{s2},W_{p2})-I(W_{s1},W_{p1};Z^n|W_{s2},W_{p2})\nonumber\\
&\quad +I(W_{s2},W_{p2};Y_2^n)-I(W_{s2},W_{p2};Z^n) +n(\gamma_n+2\epsilon_n) \\
&\leq
I(W_{s1},W_{p1};Y_1^n|W_{s2},W_{p2})-I(W_{s1},W_{p1};Z^n|W_{s2},W_{p2})+\sum_{i=1}^n
I(U_i;Y_{2i})-I(U_i;Z_i)
\nonumber\\
&\quad  +n(\gamma_n+2\epsilon_n) \label{degradedness_implies_3_implies} \\
&= \sum_{i=1}^n I(W_{s1},W_{p1};Y_{1i}|W_{s2},W_{p2},Y_1^{i-1},Z_{i+1}^n)-I(W_{s1},W_{p1};Z_i|W_{s2},W_{p2},Y_{1}^{i-1},Z_{i+1}^n)\nonumber\\
&\quad + I(U_i;Y_{2i})-I(U_i;Z_i) +n(\gamma_n+2\epsilon_n) \label{lemma_CK_2_implies_1} \\
&= \sum_{i=1}^n I(W_{s1},W_{p1};Y_{1i}|U_i)-I(W_{s1},W_{p1};Z_i|U_i) + I(U_i;Y_{2i})-I(U_i;Z_i) +n(\gamma_n+2\epsilon_n) \\
&= \sum_{i=1}^n I(W_{s1},W_{p1};Y_{1i}|U_i,Z_i) + I(U_i;Y_{2i})-I(U_i;Z_i) +n(\gamma_n+2\epsilon_n) \label{degradedness_implies_4}\\
&\leq \sum_{i=1}^n I(W_{s1},W_{p1},X_i;Y_{1i}|U_i,Z_i) + I(U_i;Y_{2i})-I(U_i;Z_i) +n(\gamma_n+2\epsilon_n) \\
&= \sum_{i=1}^n I(X_i;Y_{1i}|U_i,Z_i) + I(U_i;Y_{2i})-I(U_i;Z_i) +n(\gamma_n+2\epsilon_n) \label{memoryless}\\
&= \sum_{i=1}^n I(X_i;Y_{1i}|U_i)-I(X_i;Z_{i}|U_i) + I(U_i;Y_{2i})-I(U_i;Z_i) +n(\gamma_n+2\epsilon_n)\label{degradedness_implies_5} \\
&=\sum_{i=1}^n I(X_i;Y_{1i}|U_i)+ I(U_i;Y_{2i})-I(X_i;Z_i)
+n(\gamma_n+2\epsilon_n)\label{superposition_coding_implies}
\end{align}
where (\ref{perfect_secrecy_implies_2_implies}) comes from
(\ref{perfect_secrecy_implies_2}),
(\ref{degradedness_implies_3_implies}) is due to
(\ref{degradedness_implies_3}), (\ref{lemma_CK_2_implies_1}) comes
from Lemma~\ref{lemma_CK_2}, (\ref{degradedness_implies_4}),
(\ref{memoryless}), (\ref{degradedness_implies_5}) and
(\ref{superposition_coding_implies}) are due to the following
Markov chain
\begin{align}
W_{s1},W_{p1},U_i\rightarrow X_i \rightarrow Y_{1i} \rightarrow
Z_i
\end{align}
which is a consequence of the fact that the channel is memoryless
and degraded.

Next, we obtain an outer bound for the sum rate of the second
legitimate user's public and confidential messages $R_{p2}+R_{s2}$
as follows
\begin{align}
\lefteqn{n(R_{p2}+R_{s2})=H(W_{p2},W_{s2})}\\
&\leq I(W_{s2},W_{p2};Y_2^n)+n\epsilon_n \\
&= \sum_{i=1}^n I(W_{s2},W_{p2};Y_{2i}|Y_2^{i-1})+n\epsilon_n \\
&= \sum_{i=1}^n
I(W_{s2},W_{p2},Y_1^{i-1},Z_{i+1}^n;Y_{2i}|Y_2^{i-1})-
I(Y_1^{i-1},Z_{i+1}^n;Y_{2i}|W_{s2},W_{p2},Y_2^{i-1})+n\epsilon_n
\\
&\leq \sum_{i=1}^n
I(W_{s2},W_{p2},Y_1^{i-1},Y_2^{i-1},Z_{i+1}^n;Y_{2i})- I(Y_1^{i-1},Z_{i+1}^n;Y_{2i}|W_{s2},W_{p2},Y_2^{i-1})+n\epsilon_n\\
&= \sum_{i=1}^n
I(W_{s2},W_{p2},Y_1^{i-1},Z_{i+1}^n;Y_{2i})- I(Y_1^{i-1},Z_{i+1}^n;Y_{2i}|W_{s2},W_{p2},Y_2^{i-1})+n\epsilon_n\label{some_mc_implies}\\
&= \sum_{i=1}^n I(U_i;Y_{2i})-
I(Y_1^{i-1},Z_{i+1}^n;Y_{2i}|W_{s2},W_{p2},Y_2^{i-1})
+n\epsilon_n\label{future_use}\\
&\leq  \sum_{i=1}^n I(U_i;Y_{2i}) +n\epsilon_n
\label{partial_sum_rate_bound}
\end{align}
where (\ref{some_mc_implies}) comes from the following Markov
chain
\begin{align}
W_{s2},W_{p2},Z_{i+1}^n,Y_{2i} \rightarrow Y_1^{i-1}\rightarrow
Y_2^{i-1}
\end{align}
which is a consequence of the fact that the channel is memoryless
and degraded.

Finally, we obtain an outer bound for the sum rate
$R_{p1}+R_{s1}+R_{p2}+R_{s2}$ as follows
\begin{align}
\lefteqn{n(R_{p1}+R_{s1}+R_{p2}+R_{s2})=H(W_{p1},W_{s1},W_{p2},W_{s2})}\\
&=H(W_{p2},W_{s2})+H(W_{p1},W_{s1}|W_{p2},W_{s2})\\
&\leq \sum_{i=1}^n I(U_i;Y_{2i})
-I(Y_1^{i-1},Z_{i+1}^n;Y_{2i}|Y_2^{i-1},W_{s2},W_{p2})
+n\epsilon_n +H(W_{p1},W_{s1}|W_{p2},W_{s2})\label{future_use_implies} \\
&\leq \sum_{i=1}^n I(U_i;Y_{2i})
-I(Y_1^{i-1},Z_{i+1}^n;Y_{2i}|Y_2^{i-1},W_{s2},W_{p2})
 +I(W_{p1},W_{s1};Y_1^n|W_{p2},W_{s2})+2\epsilon_n\\
&= \sum_{i=1}^n I(U_i;Y_{2i})
-I(Y_1^{i-1},Z_{i+1}^n;Y_{2i}|Y_2^{i-1},W_{s2},W_{p2})
 +I(W_{p1},W_{s1};Y_{1i}|W_{p2},W_{s2},Y_{1}^{i-1})\nonumber\\
 &\quad +2n\epsilon_n\\
&\leq \sum_{i=1}^n I(U_i;Y_{2i})
-I(Y_1^{i-1},Z_{i+1}^n;Y_{2i}|Y_2^{i-1},W_{s2},W_{p2})
 +I(W_{p1},W_{s1},Z_{i+1}^n;Y_{1i}|W_{p2},W_{s2},Y_{1}^{i-1})\nonumber\\
 &\quad +2n\epsilon_n\\
&= \sum_{i=1}^n I(U_i;Y_{2i})
-I(Y_1^{i-1},Z_{i+1}^n;Y_{2i}|Y_2^{i-1},W_{s2},W_{p2})
 +I(Z_{i+1}^n;Y_{1i}|W_{p2},W_{s2},Y_{1}^{i-1})
 \nonumber\\
&\quad +I(W_{p1},W_{s1};Y_{1i}|W_{p2},W_{s2},Y_{1}^{i-1},Z_{i+1}^n)+2n\epsilon_n\\
&= \sum_{i=1}^n I(U_i;Y_{2i})
-I(Y_1^{i-1},Z_{i+1}^n;Y_{2i}|Y_2^{i-1},W_{s2},W_{p2})
 +I(Z_{i+1}^n;Y_{1i}|W_{p2},W_{s2},Y_{1}^{i-1})\nonumber\\
 &\quad +I(W_{p1},W_{s1};Y_{1i}|U_i)+2n\epsilon_n\\
&= \sum_{i=1}^n I(U_i;Y_{2i})
-I(Z_{i+1}^n;Y_{2i}|Y_2^{i-1},W_{s2},W_{p2})-I(Y_1^{i-1};Y_{2i}|Y_2^{i-1},W_{s2},W_{p2},Z_{i+1}^n)
\nonumber\\
&\quad +I(Z_{i+1}^n;Y_{1i}|W_{p2},W_{s2},Y_{1}^{i-1})+I(W_{p1},W_{s1};Y_{1i}|U_i)+2n\epsilon_n\\
&= \sum_{i=1}^n I(U_i;Y_{2i})
-I(Y_2^{i-1};Z_{i}|W_{s2},W_{p2},Z_{i+1}^n)
-I(Y_1^{i-1};Y_{2i}|Y_2^{i-1},W_{s2},W_{p2},Z_{i+1}^n)
\nonumber \\
&\quad+I(Y_{1}^{i-1};Z_{i}|W_{p2},W_{s2},Z_{i+1}^n)+I(W_{p1},W_{s1};Y_{1i}|U_i)+2n\epsilon_n \label{lemma_CK_1_implies}\\
&= \sum_{i=1}^n I(U_i;Y_{2i})
-I(Y_2^{i-1};Z_{i}|W_{s2},W_{p2},Z_{i+1}^n)-I(Y_1^{i-1};Y_{2i}|Y_2^{i-1},W_{s2},W_{p2},Z_{i+1}^n)
\nonumber \\
&\quad
+I(Y_{2}^{i-1},Y_{1}^{i-1};Z_{i}|W_{p2},W_{s2},Z_{i+1}^n)+I(W_{p1},W_{s1};Y_{1i}|U_i)+2n\epsilon_n
\label{degradedness_implies_6}
\end{align}
\begin{align}
&= \sum_{i=1}^n
I(U_i;Y_{2i})-I(Y_1^{i-1};Y_{2i}|Y_2^{i-1},W_{s2},W_{p2},Z_{i+1}^n)+I(Y_{1}^{i-1};Z_{i}|W_{p2},W_{s2},Z_{i+1}^n,Y_{2}^{i-1})
\nonumber\\
&\quad +I(W_{p1},W_{s1};Y_{1i}|U_i)+2n\epsilon_n \\
&= \sum_{i=1}^n
I(U_i;Y_{2i})-I(Y_1^{i-1};Y_{2i},Z_i|Y_2^{i-1},W_{s2},W_{p2},Z_{i+1}^n)
+I(Y_{1}^{i-1};Z_{i}|W_{p2},W_{s2},Z_{i+1}^n,Y_{2}^{i-1})
\nonumber\\
&\quad +I(W_{p1},W_{s1};Y_{1i}|U_i)+2n\epsilon_n \label{degradedness_implies_7} \\
&= \sum_{i=1}^n I(U_i;Y_{2i})-I(Y_1^{i-1};Y_{2i}|Y_2^{i-1},W_{s2},W_{p2},Z_{i+1}^n,Z_i) +I(W_{p1},W_{s1};Y_{1i}|U_i)+2n\epsilon_n  \\
&\leq \sum_{i=1}^n I(U_i;Y_{2i})+I(W_{p1},W_{s1};Y_{1i}|U_i)+2n\epsilon_n \\
&\leq \sum_{i=1}^n I(U_i;Y_{2i})+I(W_{p1},W_{s1},X_i;Y_{1i}|U_i)+2n\epsilon_n \\
&= \sum_{i=1}^n I(U_i;Y_{2i})+I(X_i;Y_{1i}|U_i)+2n\epsilon_n
\label{memoryless_1}
\end{align}
where (\ref{future_use_implies}) comes from (\ref{future_use}),
(\ref{lemma_CK_1_implies}) is obtained by using
Lemma~\ref{lemma_CK_1}, (\ref{degradedness_implies_6}) is due to
the following Markov chain
\begin{align}
Y_2^{i-1} \rightarrow Y_1^{i-1} \rightarrow
W_{p2},W_{s2},Z_{i+1}^n,Z_i
\end{align}
which comes from the fact that the channel is degraded and
memoryless, (\ref{degradedness_implies_7}) is a consequence of the
following Markov chain
\begin{align}
W_{s2}W_{p2}Y_1^{i-1}Y_2^{i-1}Z_{i+1}^n \rightarrow Y_{2i}
\rightarrow Z_i
\end{align}
which also comes from the fact that the channel is degraded and
memoryless, and (\ref{memoryless_1}) is due to the following
Markov chain
\begin{align}
W_{s1}W_{p1}U_i \rightarrow X_i \rightarrow Y_{1i}
\end{align}
which is a consequence of the fact that the channel is memoryless.
The bounds in (\ref{degradedness_implies_3}),
(\ref{superposition_coding_implies}),
(\ref{partial_sum_rate_bound}) and (\ref{memoryless_1}) can be
single-letterized yielding the bounds given in
Theorem~\ref{theorem_outer_bound}.

\section{Proof of Theorem~\ref{theorem_inner_bound_general}}
\label{proof_of_theorem_inner_bound_general}

Here, we first consider a more general scenario than the scenario
introduced in Section~\ref{sec:inner_general}. In this more
general scenario, the transmitter sends a pair of common public
and confidential messages to the legitimate users in addition to a
pair of public and confidential messages intended to each
legitimate user. Thus, in this case, the transmitter has the
message tuple $(W_{p0},W_{s0},W_{p1},W_{s1},W_{p2},W_{s2})$, where
the common public message $W_{p0}$ and the common confidential
message $W_{s0}$ are sent to both legitimate users, and a pair of
public and confidential messages $(W_{pj},W_{sj})$ are sent to the
$j$th legitimate user,$~j=1,2$\footnote{Another way to obtain the
achievable rate region in
Theorem~\ref{theorem_inner_bound_general} is to use rate-splitting
for $W_{p1},W_{s1},W_{p2},W_{s2}$ as we mentioned earlier in
Section~\ref{sec:inner_general}. However, we chose to introduce a
pair of common public and confidential messages here, because the
corresponding scenario results in an achievable scheme that
encompasses the one we can obtain by using rate-splitting.}. Here
also, there is no secrecy concern on the public messages
$W_{p0},W_{p1},W_{p2}$ while the confidential messages
$W_{s0},W_{s1},W_{s2}$ need to be transmitted in perfect secrecy,
i.e., they need to satisfy the following
\begin{align}
\lim_{n\rightarrow \infty} \frac{1}{n}
I(W_{s0},W_{s1},W_{s2};Z^n)=0 \label{perfect_secrecy_general}
\end{align}
The definition of a code, achievable rates, and the capacity
region for this more general scenario can be given similar to the
definitions of a code, achievable rates, and the capacity region
we provided in Section~\ref{sec:discrete_model}. Here, we provide
the proof of Theorem~\ref{theorem_inner_bound_general} in two
steps. In the first step, we prove an achievable rate region for
the more general scenario we just introduced. In the second step,
we obtain the achievable rate region in
Theorem~\ref{theorem_inner_bound_general} from the achievable rate
region proved in the first step by using Fourier-Motzkin
elimination in conjunction with the fact that since the common
public and confidential messages $W_{p0},W_{s0}$ are decoded by
both users, they can be converted into public and confidential
messages $(W_{p1}, W_{s1},W_{p2},W_{s2})$ of the legitimate users.

\subsection{Part I}
We fix the joint distribution $p(q,u,v_1,v_2,x,y_1,y_2,z)$ as
follows
\begin{align}
p(q,u)p(v_1,v_2,x|u)p(y_1,y_2,z|x)
\end{align}
Next, we divide the common public message rate $R_{p0}$ into two
parts as follows.
\begin{align}
R_{p0}&=\tilde{R}_{p0}+\dtildeR_{p0}
\end{align}
In other words, we divide the common public message $W_{p0}$ into
two parts as $W_{p0}=(\tilde{W}_{p0},\tilde{\tilde{W}}_{p0})$,
where the rate of $\tilde{W}_{p0}$ is $\tilde{R}_{p0}$, and the
rate of $\tilde{\tilde{W}}_{p0}$ is $\tilde{\tilde{R}}_{p0}$. We
use rate-splitting for the common public message because due
to~\cite{Korner}, we know that rate-splitting might enhance the
achievable public and confidential message rate pairs even for the
single legitimate user case.

\vspace{0.25cm} \noindent \underline{Codebook generation:}
\begin{itemize}
\item Generate $2^{n\tilde{R}_{p0}}$ length-$n$ sequences $q^n$
 through $p(q^n)=\prod_{i=1}^np(q_i)$, and index them as
 $q^{n}(\tilde{w}_{p0})$,
 where $\tilde{w}_{p0}\in\{1,\ldots,2^{n\tilde{R}_{p0}}\}$.

 \item For each $q^{n}(\tilde{w}_{p0})$ sequence, generate $2^{n(\dtildeR_{p0}+R_{s0}+\Delta_0)}$ length-$n$ sequences $u^n$
 through $p(u^n|q^n)=\prod_{i=1}^np(u_i|q_i)$, and index them as
 $u^{n}(\tilde{w}_{p0},\dtildew_{p0},w_{s0},d_0)$,
 where $\dtildew_{p0}\in\{1,\ldots,\break 2^{n\dtildeR_{p0}}\}, w_{s0}\in\{1,\ldots,2^{nR_{s0}}\},d_0\in\{1,\ldots,2^{n\Delta_0}\}$.

 \item For each $u^{n}(\tilde{w}_{p0},\dtildew_{p0},w_{s0},d_0)$ sequence, generate $2^{n(R_{p1}+R_{s1}+\Delta_1+L_1)}$ length-$n$ sequences $v_1^n$
 through $p(v_1^n|u^n)=\prod_{i=1}^np(v_{1i}|u_i)$, and index them as $v_1^{n}(\tilde{w}_{p0},\dtildew_{p0},w_{s0},d_0,w_{p1},w_{s1},d_1,\break
 l_1)$,
 where $w_{p1}\in\{1,\ldots,2^{nR_{p1}}\},w_{s1}\in\{1,\ldots,2^{nR_{s1}}\},d_1\in\{1,\ldots,2^{n\Delta_{1}}\},l_1\in\{1,\ldots,2^{nL_1}\}$.

 \item For each $u^{n}(\tilde{w}_{p0},\dtildew_{p0},w_{s0},d_0)$ sequence, generate $2^{n(R_{p2}+R_{s2}+\Delta_2+L_2)}$ length-$n$ sequences $v_2^n$
 through $p(v_2^n|u^n)=\prod_{i=1}^np(v_{2i}|u_i)$, and index them as $v_2^{n}(\tilde{w}_{p0},\dtildew_{p0},w_{s0},d_0,w_{p2},w_{s2},d_2,\break
 l_2)$,
 where $w_{p2}\in\{1,\ldots,2^{nR_{p2}}\},w_{s2}\in\{1,\ldots,2^{nR_{s2}}\},d_2\in\{1,\ldots,2^{n\Delta_{2}}\},l_2\in\{1,\ldots,2^{nL_2}\}$.

 \end{itemize}

\vspace{0.25cm} \noindent \underline{Encoding:} Assume
$W_{p0}=w_{p0},W_{s0}=w_{s0},W_{p1}=w_{p1},W_{s1}=w_{s1},W_{p2}=w_{p2},W_{s2}=w_{s2}$
is the message to be transmitted. Randomly pick $d_0,d_1,d_2$.
Next, we find an $(l_1,l_2)$ pair such that the corresponding
sequence tuple $(q^n,u^n,v_1^n,v_2^n)$ is jointly typical. Due to
mutual covering lemma~\cite{Abbas-El}, if $L_1,L_2$ satisfy
\begin{align}
L_1+L_2 \geq I(V_1;V_2|U) \label{ach_gen_constraint_1}
\end{align}
with high probability, there exists at least one pair $(l_1,l_2)$
such that the corresponding sequence tuple $(q^n,u^n,v_1^n,v_2^n)$
is jointly typical.

\vspace{0.25cm}\noindent \underline{Decoding:}

\begin{itemize}
\item The first legitimate user decodes
$(w_{p0},w_{s0},d_0,w_{p1},w_{s1},d_1)$ in two steps. In the first
step, it decodes $(w_{p0},w_{s0},d_0)$ by looking for the unique
$(q^n,u^n)$ pair such that $ (q^n,u^n,y_1^n)$ is jointly typical.
If the following conditions are satisfied,
\begin{align}
R_{p0}+R_{s0}+\Delta_0 &\leq I(U;Y_1)\label{ach_gen_constraint_2}\\
\tilde{\tilde{R}}_{p0}+R_{s0}+\Delta_0 &\leq
I(U;Y_1|Q)\label{ach_gen_constraint_3}
\end{align}
the first legitimate user can decode $(w_{p0},w_{s0},d_0)$ with
vanishingly small probability of error.

In the second step, it decodes $(w_{s1},w_{p1},d_1)$ by looking
for the unique $(q^n,u^n,v_1^n)$ tuple such that $
(q^n,u^n,v_1^n,y_1^n)$ is jointly typical. Assuming that
$(w_{p0},w_{s0},d_0)$ is decoded correctly in the first step, if
the following condition is satisfied,
\begin{align}
R_{p1}+R_{s1}+\Delta_1+L_1 &\leq I(V_1;Y_1|U)
\label{ach_gen_constraint_4}
\end{align}
the first legitimate user can decode $(w_{p1},w_{s1},d_1)$ with
vanishingly small probability of error.

\item Similarly, the second legitimate user can decode
$(w_{p0},w_{s0},d_0,w_{p2},w_{s2},d_2)$ with vanishingly small
probability of error if the following conditions are satisfied.
\begin{align}
R_{p0}+R_{s0}+\Delta_0 &\leq
I(U;Y_2) \label{ach_gen_constraint_5}\\
\tilde{\tilde{R}}_{p0}+R_{s0}+\Delta_0 &\leq
I(U;Y_2|Q)\label{ach_gen_constraint_6}\\
R_{p2}+R_{s2}+\Delta_2+L_2 &\leq I(V_2;Y_2|U)
\label{ach_gen_constraint_7}
\end{align}

\end{itemize}

\noindent \underline{Equivocation computation:} We now show that
the proposed coding scheme satisfies the perfect secrecy
requirement on the confidential messages given by
(\ref{perfect_secrecy_general}). We start as follows.
\begin{align}
\lefteqn{H(W_{s0},W_{s1},W_{s2}|Z^n)}\nonumber\\
&\geq  H(W_{s0},W_{s1},W_{s2}|Z^n,Q^n)\\
&=
H(W_{s0},W_{s1},W_{s2},\tilde{\tilde{W}}_{p0},W_{p1},W_{p2},D_0,D_1,D_2|Z^n,Q^n)
\nonumber\\
&\quad
-H(\tilde{\tilde{W}}_{p0},W_{p1},W_{p2},D_0,D_1,D_2|Z^n,Q^n,W_{s0},W_{s1},W_{s2})\\
&=
H(W_{s0},W_{s1},W_{s2},\tilde{\tilde{W}}_{p0},W_{p1},W_{p2},D_0,D_1,D_2|Q^n)\nonumber\\
&\quad
-I(W_{s0},W_{s1},W_{s2},\tilde{\tilde{W}}_{p0},W_{p1},W_{p2},D_0,D_1,D_2;Z^n|Q^n)
\nonumber\\
&\quad
-H(\tilde{\tilde{W}}_{p0},W_{p1},W_{p2},D_0,D_1,D_2|Z^n,Q^n,W_{s0},W_{s1},W_{s2})\\
&= H(W_{s0},W_{s1},W_{s2})
+H(\tilde{\tilde{W}}_{p0},W_{p1},W_{p2},D_0,D_1,D_2)\nonumber\\
&\quad
-I(W_{s0},W_{s1},W_{s2},\tilde{\tilde{W}}_{p0},W_{p1},W_{p2},D_0,D_1,D_2;Z^n|Q^n)
\nonumber\\
&\quad
-H(\tilde{\tilde{W}}_{p0},W_{p1},W_{p2},D_0,D_1,D_2|Z^n,Q^n,W_{s0},W_{s1},W_{s2})\label{independence_of_messages}\\
&= H(W_{s0},W_{s1},W_{s2})
+n(\tilde{\tilde{R}}_{p0}+R_{p1}+R_{p2}+\Delta_0+\Delta_1+\Delta_2)\nonumber\\
&\quad
-I(W_{s0},W_{s1},W_{s2},\tilde{\tilde{W}}_{p0},W_{p1},W_{p2},D_0,D_1,D_2;Z^n|Q^n)
\nonumber\\
&\quad
-H(\tilde{\tilde{W}}_{p0},W_{p1},W_{p2},D_0,D_1,D_2|Z^n,Q^n,W_{s0},W_{s1},W_{s2})\label{independence_of_messages_again}\\
&\geq  H(W_{s0},W_{s1},W_{s2})
+n(\tilde{\tilde{R}}_{p0}+R_{p1}+R_{p2}+\Delta_0+\Delta_1+\Delta_2)\nonumber\\
&\quad
-I(W_{s0},W_{s1},W_{s2},\tilde{\tilde{W}}_{p0},W_{p1},W_{p2},D_0,D_1,D_2,U^n,V_1^n,V_2^n;Z^n|Q^n)
\nonumber\\
&\quad
-H(\tilde{\tilde{W}}_{p0},W_{p1},W_{p2},D_0,D_1,D_2|Z^n,Q^n,W_{s0},W_{s1},W_{s2})\\
&=  H(W_{s0},W_{s1},W_{s2})
+n(\tilde{\tilde{R}}_{p0}+R_{p1}+R_{p2}+\Delta_0+\Delta_1+\Delta_2)\nonumber\\
&\quad -I(U^n,V_1^n,V_2^n;Z^n|Q^n)
-H(\tilde{\tilde{W}}_{p0},W_{p1},W_{p2},D_0,D_1,D_2|Z^n,Q^n,W_{s0},W_{s1},W_{s2})\label{messages_and_codewords}\\
&\geq   H(W_{s0},W_{s1},W_{s2})
+n(\tilde{\tilde{R}}_{p0}+R_{p1}+R_{p2}+\Delta_0+\Delta_1+\Delta_2)\nonumber\\
&\quad -n(I(U,V_1,V_2;Z|Q)+\gamma_{1n})
-H(\tilde{\tilde{W}}_{p0},W_{p1},W_{p2},D_0,D_1,D_2|Z^n,Q^n,W_{s0},W_{s1},W_{s2})
\label{ruoheng_showed_it}
\end{align}
where
(\ref{independence_of_messages})-(\ref{independence_of_messages_again})
follow from the facts that the messages
$W_{s0},W_{s1},W_{s2},\tilde{\tilde{W}}_{p0},W_{p1},W_{p2},D_0,D_1,\break
D_2$ are independent among themselves, uniformly distributed, and
also are independent of $Q^n$, (\ref{messages_and_codewords})
stems from the fact that given the codewords
$(Q^n,U^n,V_1^n,V_2^n)$,
$(W_{s0},W_{s1},W_{s2},\tilde{\tilde{W}}_{p0},\break
W_{p1},W_{p2},D_0,D_1,D_2)$ and $Z^n$ are independent,
(\ref{ruoheng_showed_it}) comes from the fact that
\begin{align}
I(U^n,V_1^n,V_2^n;Z^n|Q^n)\leq n I(U,V_1,V_2;Z|Q)+n\gamma_{1n}
\label{what_ruoheng_showed}
\end{align}
where $\gamma_{1n}\rightarrow 0$ as $n\rightarrow \infty$. The
bound in (\ref{what_ruoheng_showed}) can be shown by following the
analysis in~\cite{ruoheng}. Next, we consider the conditional
entropy term in (\ref{ruoheng_showed_it}). To this end, we
introduce the following lemma.
\begin{Lem}
\label{lemma_eavesdropper_can_decode} We have
\begin{align}
H(W_{p1},W_{p2},D_1,D_2|Z^n,Q^n,W_{s0},W_{s1},W_{s2},\tilde{\tilde{W}}_{p0},D_0)\leq
n\gamma_{2n}
\end{align}
where $\gamma_{2n}\rightarrow 0$ as $n\rightarrow \infty$, if the
following conditions are satisfied.
\begin{align}
R_{p1}+\Delta_1+L_1+R_{p2}+\Delta_2+L_2
&\leq I(V_1,V_2;Z|U)+I(V_1;V_2|U)\label{decodability_at_eavesdropper_1}\\
R_{p1}+\Delta_1+L_1
&\leq I(V_1;Z,V_2|U)\label{decodability_at_eavesdropper_2}\\
R_{p2}+\Delta_2+L_2 &\leq I(V_2;Z,V_1|U)
\label{decodability_at_eavesdropper_3}
\end{align}
\end{Lem}
The proof of Lemma~\ref{lemma_eavesdropper_can_decode} is given in
Appendix~\ref{proof_eavesdropper_can_decode}. Using this lemma, we
have the following corollary.
\begin{Cor}
\label{corollary_eavesdropper_can_decode} We have
\begin{align}
H(\tilde{\tilde{W}}_{p0},D_0|Z^n,Q^n,W_{s0},W_{s1},W_{s2})\leq
n\gamma_{3n}
\end{align}
where $\gamma_{3n}\rightarrow 0$ as $n\rightarrow \infty$, if the
following condition is satisfied.
\begin{align}
\tilde{\tilde{R}}_{p0}+\Delta_0 &\leq I(U;Z|Q)
\label{decodability_at_eavesdropper_4}
\end{align}
\end{Cor}
Now, we set the rates
$\tilde{\tilde{R}}_{p0},\Delta_0,R_{p1},\Delta_1,L_1,R_{p2},\Delta_2,
L_2$ as follows.
\begin{align}
\tilde{\tilde{R}}_{p0}+\Delta_0&=I(U;Z|Q)-\epsilon \label{ach_gen_constraint_8}\\
L_1+L_2&=I(V_1;V_2|U)+\frac{\epsilon}{2} \label{ach_gen_constraint_9} \\
R_{p1}+\Delta_1+R_{p2}+\Delta_2&=I(V_1,V_2;Z|U)-\epsilon \label{ach_gen_constraint_10}\\
R_{p1}+\Delta_1+L_1 &< I(V_1;Z,V_2|U) \label{ach_gen_constraint_11}\\
R_{p2}+\Delta_2+L_2 &< I(V_2;Z,V_1|U)
\label{ach_gen_constraint_12}
\end{align}
In view of Lemma~\ref{lemma_eavesdropper_can_decode} and
Corollary~\ref{corollary_eavesdropper_can_decode}, the selections
of
$\tilde{\tilde{R}}_{p0},\Delta_0,R_{p1},\Delta_1,L_1,R_{p2},\Delta_2,L_2$
in (\ref{ach_gen_constraint_8})-(\ref{ach_gen_constraint_12})
imply that
\begin{align}
\lefteqn{H(\tilde{\tilde{W}}_{p0},W_{p1},W_{p2},D_0,D_1,D_2|Z^n,Q^n,W_{s0},W_{s1},W_{s2})}\nonumber\\
&=H(\tilde{\tilde{W}}_{p0},D_0|Z^n,Q^n,W_{s0},W_{s1},W_{s2})
+H(W_{p1},W_{p2},D_1,D_2|Z^n,Q^n,W_{s0},W_{s1},W_{s2},\tilde{\tilde{W}}_{p0},D_0)\\
&\leq n(\gamma_{2n}+\gamma_{3n})
\end{align}
using which in (\ref{ruoheng_showed_it}), we get
\begin{align}
H(W_{s0},W_{s1},W_{s2}|Z^n)&\geq H(W_{s0},W_{s1},W_{s2})
+n(\tilde{\tilde{R}}_{p0}+R_{p1}+R_{p2}+\Delta_0+\Delta_1+\Delta_2)\nonumber\\
&\quad -n(I(U,V_1,V_2;Z|Q)+\gamma_{1n})
-n(\gamma_{2n}+\gamma_{3n}) \label{equivocation_almost_done}
\end{align}
Moreover, using
(\ref{ach_gen_constraint_8})-(\ref{ach_gen_constraint_10}) in
(\ref{equivocation_almost_done}), we have
\begin{align}
H(W_{s0},W_{s1},W_{s2}|Z^n)&\geq H(W_{s0},W_{s1},W_{s2})
-n\frac{3\epsilon}{2}-n (\gamma_{1n}+\gamma_{2n}+\gamma_{3n})
\end{align}
which implies that the proposed coding scheme satisfies the
perfect secrecy requirement on the confidential messages;
completing the equivocation computation.

\subsection{Part II}

\label{sec:Fourier-Motzkin} We have shown that rate tuples
$(R_{p0},R_{s0},R_{p1},R_{s1},R_{p2},R_{s2})$ satisfying the
following constraints are achievable.
\begin{align}
L_1+L_2 &= I(V_1;V_2|U) \label{ach_gen_again_v1_first}\\
R_{p0}+R_{s0}+\Delta_0 &\leq \min_{j=1,2}I(U;Y_j)\\
\tilde{\tilde{R}}_{p0}+R_{s0}+\Delta_0 &\leq \min_{j=1,2}I(U;Y_j|Q)\\
R_{p1}+R_{s1}+\Delta_1+L_1 &\leq I(V_1;Y_1|U)\\
R_{p2}+R_{s2}+\Delta_2+L_2 &\leq I(V_2;Y_2|U) \\
\tilde{\tilde{R}}_{p0}+\Delta_0&=I(U;Z|Q) \\
R_{p1}+\Delta_1+R_{p2}+\Delta_2&=I(V_1,V_2;Z|U) \\
R_{p1}+\Delta_1+L_1 &\leq I(V_1;Z,V_2|U) \\
R_{p2}+\Delta_2+L_2 &\leq I(V_2;Z,V_1|U)
\label{ach_gen_again_v1_last}
\end{align}
We eliminate $\Delta_0$ from the bounds in
(\ref{ach_gen_again_v1_first})-(\ref{ach_gen_again_v1_last}) by
using Fourier-Motzkin elimination, which leads to the following
region.
\begin{align}
L_1+L_2 &= I(V_1;V_2|U) \label{ach_gen_again_v2_first}\\
R_{p0}+R_{s0}+I(U;Z|Q)-\tilde{\tilde{R}}_{p0} &\leq \min_{j=1,2}I(U;Y_j)\\
R_{s0} &\leq \min_{j=1,2}I(U;Y_j|Q)-I(U;Z|Q)\\
R_{p1}+R_{s1}+\Delta_1+L_1 &\leq I(V_1;Y_1|U)\\
R_{p2}+R_{s2}+\Delta_2+L_2 &\leq I(V_2;Y_2|U) \\
R_{p1}+\Delta_1+R_{p2}+\Delta_2&=I(V_1,V_2;Z|U) \\
R_{p1}+\Delta_1+L_1 &\leq I(V_1;Z,V_2|U) \\
R_{p2}+\Delta_2+L_2 &\leq I(V_2;Z,V_1|U)\\
0&\leq \tilde{\tilde{R}}_{p0}\\
\tilde{\tilde{R}}_{p0}&\leq R_{p0}\\
\tilde{\tilde{R}}_{p0}&\leq I(U;Z|Q) \label{ach_gen_again_v2_last}
\end{align}
Next, we eliminate $\tilde{\tilde{R}}_{p0}$ from the bounds in
(\ref{ach_gen_again_v2_first})-(\ref{ach_gen_again_v2_last}) by
using Fourier-Motzkin elimination, which leads to the following
region.
\begin{align}
L_1+L_2 &= I(V_1;V_2|U) \label{ach_gen_again_v3_first}\\
R_{p0}+R_{s0} &\leq \min_{j=1,2}I(U;Y_j)\\
R_{s0} &\leq \min_{j=1,2}I(U;Y_j|Q)-I(U;Z|Q)\\
R_{p1}+R_{s1}+\Delta_1+L_1 &\leq I(V_1;Y_1|U)\\
R_{p2}+R_{s2}+\Delta_2+L_2 &\leq I(V_2;Y_2|U) \\
R_{p1}+\Delta_1+R_{p2}+\Delta_2&=I(V_1,V_2;Z|U) \\
R_{p1}+\Delta_1+L_1 &\leq I(V_1;Z,V_2|U) \\
R_{p2}+\Delta_2+L_2 &\leq I(V_2;Z,V_1|U)\\
\label{ach_gen_again_v3_last}
\end{align}
Next, we eliminate $L_1$ from the bounds in
(\ref{ach_gen_again_v3_first})-(\ref{ach_gen_again_v3_last}) by
using Fourier-Motzkin elimination, which leads to the following
region.
\begin{align}
R_{p0}+R_{s0} &\leq \min_{j=1,2}I(U;Y_j) \label{ach_gen_again_v4_first}\\
R_{s0} &\leq \min_{j=1,2}I(U;Y_j|Q)-I(U;Z|Q)\\
R_{p1}+R_{s1}+\Delta_1+I(V_1;V_2|U)-L_2  &\leq I(V_1;Y_1|U)\\
R_{p2}+R_{s2}+\Delta_2+L_2 &\leq I(V_2;Y_2|U) \\
R_{p1}+\Delta_1+R_{p2}+\Delta_2&=I(V_1,V_2;Z|U) \\
R_{p1}+\Delta_1+I(V_1;V_2|U)-L_2  &\leq I(V_1;Z,V_2|U) \\
R_{p2}+\Delta_2+L_2 &\leq I(V_2;Z,V_1|U)\\
L_2&\leq I(V_1;V_2|U) \\
0&\leq L_2 \label{ach_gen_again_v4_last}
\end{align}
Next, we eliminate $L_2$ from the bounds in
(\ref{ach_gen_again_v4_first})-(\ref{ach_gen_again_v4_last}) by
using Fourier-Motzkin elimination, which leads to the following
region.
\begin{align}
R_{p0}+R_{s0} &\leq \min_{j=1,2}I(U;Y_j) \label{ach_gen_again_v5_first}\\
R_{s0} &\leq \min_{j=1,2}I(U;Y_j|Q)-I(U;Z|Q)\\
R_{s1}&\leq I(V_1;Y_1|U)-I(V_1;Z|U)\\
R_{s2}&\leq I(V_2;Y_2|U)-I(V_2;Z|U)\\
R_{s1}+R_{s2}&\leq
I(V_1;Y_1|U)+I(V_2;Y_2|U)-I(V_1;V_2|U)-I(V_1,V_2;Z|U)\\
R_{p1}+R_{s1}+\Delta_1&\leq I(V_1;Y_1|U)\\
R_{p2}+R_{s2}+\Delta_2&\leq I(V_2;Y_2|U)\\
R_{p1}+\Delta_1&\leq I(V_1;Z,V_2|U)\\
R_{p2}+\Delta_2&\leq I(V_2;Z,V_1|U)\\
R_{p1}+\Delta_1+R_{p2}+\Delta_2&=I(V_1,V_2;Z|U)
\label{ach_gen_again_v5_last}
\end{align}
We eliminate $\Delta_1$ from the bounds in
(\ref{ach_gen_again_v5_first})-(\ref{ach_gen_again_v5_last}) by
using Fourier-Motzkin elimination, which leads to the following
region.
\begin{align}
R_{p0}+R_{s0} &\leq \min_{j=1,2}I(U;Y_j) \label{ach_gen_again_v6_first}\\
R_{s0} &\leq \min_{j=1,2}I(U;Y_j|Q)-I(U;Z|Q)\\
R_{s1}&\leq I(V_1;Y_1|U)-I(V_1;Z|U)\\
R_{s2}&\leq I(V_2;Y_2|U)-I(V_2;Z|U)\\
R_{s1}+R_{s2}&\leq
I(V_1;Y_1|U)+I(V_2;Y_2|U)-I(V_1;V_2|U)\nonumber\\
&\quad -I(V_1,V_2;Z|U)\\
R_{s1}+I(V_1,V_2;Z|U)-R_{p2}-\Delta_2&\leq I(V_1;Y_1|U)\\
R_{p2}+R_{s2}+\Delta_2&\leq I(V_2;Y_2|U)\\
I(V_1,V_2;Z|U)-R_{p2}-\Delta_2 &\leq I(V_1;Z,V_2|U)\\
R_{p2}+\Delta_2&\leq I(V_2;Z,V_1|U)\\
0&\leq I(V_1,V_2;Z|U)-R_{p1}-R_{p2}-\Delta_2 \\
0&\leq \Delta_2 \label{ach_gen_again_v6_last}
\end{align}
We eliminate $\Delta_2$ from the bounds in
(\ref{ach_gen_again_v6_first})-(\ref{ach_gen_again_v6_last}) by
using Fourier-Motzkin elimination, which leads to the following
region.
\begin{align}
R_{p0}+R_{s0} &\leq \min_{j=1,2}I(U;Y_j) \label{ach_gen_again_v7_first}\\
R_{s0} &\leq \min_{j=1,2}I(U;Y_j|Q)-I(U;Z|Q)\\
R_{s1}&\leq I(V_1;Y_1|U)-I(V_1;Z|U)\\
R_{s2}&\leq I(V_2;Y_2|U)-I(V_2;Z|U)\\
R_{s1}+R_{s2}&\leq
I(V_1;Y_1|U)+I(V_2;Y_2|U)-I(V_1;V_2|U)-I(V_1,V_2;Z|U)\\
R_{p1}+R_{s1}&\leq I(V_1;Y_1|U) \\
R_{p2}+R_{s2}&\leq I(V_2;Y_2|U) \\
R_{p1}&\leq I(V_1;Z,V_2|U)\\
R_{p2}&\leq I(V_2;Z,V_1|U)\\
R_{p1}+R_{p2}&\leq I(V_1,V_2;Z|U)
 \label{ach_gen_again_v7_last}
\end{align}
We note the fact that each legitimate user's own confidential
message rate $R_{sj}$ can be given up in favor of its public
message rate $R_{pj},~j=1,2$, i.e., if the rate tuple
$(R_{p0},R_{s0},R_{p1},R_{s1},R_{p2},R_{s2})$ is achievable, the
rate tuple
$(R_{p0},R_{s0},R_{p1}+\alpha_1,R_{s1}-\alpha_1,R_{p2}+\alpha_2,R_{s2}-\alpha_2)$
is also achievable for any non-negative $(\alpha_1,\alpha_2)$
pairs satisfying $\alpha_1\leq R_{s1},\alpha_2\leq R_{s2}$. Using
this fact for the region given by
(\ref{ach_gen_again_v7_first})-(\ref{ach_gen_again_v7_last}), we
obtain the following region.
\begin{align}
R_{p0}+R_{s0} &\leq \min_{j=1,2}I(U;Y_j) \label{ach_gen_again_v8_first}\\
R_{s0} &\leq \min_{j=1,2}I(U;Y_j|Q)-I(U;Z|Q)\\
R_{s1}+\alpha_1&\leq I(V_1;Y_1|U)-I(V_1;Z|U)\\
R_{s2}+\alpha_2&\leq I(V_2;Y_2|U)-I(V_2;Z|U)\\
R_{s1}+R_{s2}+\alpha_1+\alpha_2&\leq
I(V_1;Y_1|U)+I(V_2;Y_2|U)-I(V_1;V_2|U)-I(V_1,V_2;Z|U)\\
R_{p1}+R_{s1}&\leq I(V_1;Y_1|U) \\
R_{p2}+R_{s2}&\leq I(V_2;Y_2|U) \\
R_{p1}-\alpha_1&\leq I(V_1;Z,V_2|U)\\
R_{p2}-\alpha_2&\leq I(V_2;Z,V_1|U)\\
R_{p1}+R_{p2}-\alpha_1-\alpha_2&\leq I(V_1,V_2;Z|U)\\
0&\leq \alpha_1\\
0&\leq \alpha_2\\
\alpha_1&\leq R_{p1}\\
\alpha_2 &\leq R_{p2}
 \label{ach_gen_again_v8_last}
\end{align}
We first eliminate $\alpha_1$ from the bounds in
(\ref{ach_gen_again_v8_first})-(\ref{ach_gen_again_v8_last}) by
using Fourier-Motzkin elimination, which leads to the following
region.
\begin{align}
R_{p0}+R_{s0} &\leq \min_{j=1,2}I(U;Y_j) \label{ach_gen_again_v9_first}\\
R_{s0} &\leq \min_{j=1,2}I(U;Y_j|Q)-I(U;Z|Q)\\
R_{s1}&\leq I(V_1;Y_1|U)-I(V_1;Z|U)\\
R_{s1}+R_{p1}+R_{p2}-\alpha_2&\leq
I(V_1;Y_1|U)-I(V_1;Z|U)+I(V_1,V_2;Z|U)\\
R_{s1}+R_{p1}+R_{s2}+\alpha_2&\leq
I(V_1;Y_1|U)+I(V_2;Y_2|U)-I(V_2;Z|U)\\
R_{s1}+R_{p1}+R_{s2}+R_{p2}&\leq
I(V_1;Y_1|U)+I(V_2;Y_2|U)-I(V_1;V_2|U)\\
R_{s1}+R_{s2}+\alpha_2 &\leq
I(V_1;Y_1|U)+I(V_2;Y_2|U)-I(V_1;V_2|U)-I(V_1,V_2;Z|U)\\
R_{p2}-\alpha_2&\leq I(V_1,V_2;Z|U)\\
R_{s2}+\alpha_2&\leq I(V_2;Y_2|U)-I(V_2;Z|U)\\
R_{p1}+R_{s1}&\leq I(V_1;Y_1|U) \\
R_{p2}+R_{s2}&\leq I(V_2;Y_2|U) \\
R_{p2}-\alpha_2&\leq I(V_2;Z,V_1|U)\\
0&\leq \alpha_2\\
\alpha_2 &\leq R_{p2}
 \label{ach_gen_again_v9_last}
\end{align}
Next, we eliminate $\alpha_2$ from the bounds in
(\ref{ach_gen_again_v9_first})-(\ref{ach_gen_again_v9_last}) by
using Fourier-Motzkin elimination, which leads to the following
region.
\begin{align}
R_{p0}+R_{s0} &\leq \min_{j=1,2}I(U;Y_j) \label{ach_gen_again_v10_first}\\
R_{s0} &\leq \min_{j=1,2}I(U;Y_j|Q)-I(U;Z|Q)\\
R_{s1}&\leq I(V_1;Y_1|U)-I(V_1;Z|U)\\
R_{s2}&\leq I(V_2;Y_2|U)-I(V_2;Z|U)\\
R_{s1}+R_{s2}&\leq I(V_1;Y_1|U)+I(V_2;Y_2|U)-I(V_1;V_2|U)
-I(V_1,V_2;Z|U)\\
R_{p1}+R_{s1}&\leq I(V_1;Y_1|U) \\
R_{p2}+R_{s2}&\leq I(V_2;Y_2|U)\\
R_{s1}+R_{p1}+R_{s2}&\leq I(V_1;Y_1|U)+I(V_2;Y_2|U)-I(V_2;Z|U)\\
R_{s1}+R_{s2}+R_{p2}&\leq I(V_1;Y_1|U)+I(V_2;Y_2|U)-I(V_1;Z|U)\\
R_{s1}+R_{p1}+R_{s2}+R_{p2}&\leq
I(V_1;Y_1|U)+I(V_2;Y_2|U)-I(V_1;V_2|U)
 \label{ach_gen_again_v10_last}
\end{align}
We note the fact that since the common public message is decoded
by both legitimate users, its rate $R_{p0}$ can be given up in
favor of each legitimate user's own public message rate
$R_{pj},~j=1,2$, i.e., if the rate tuple
$(R_{s0},R_{p0},R_{s1},R_{p1},R_{s2},R_{p2})$ is achievable, the
rate tuple
$(R_{p0}-\alpha-\beta,R_{s0},R_{p1}+\alpha,R_{s1},R_{p2}+\beta,R_{s2})$
is also achievable for any non-negative $(\alpha,\beta)$ pairs
satisfying $\alpha+\beta\leq R_{p0}$. Using this fact for the
region given by
(\ref{ach_gen_again_v10_first})-(\ref{ach_gen_again_v10_last}), we
obtain the following region.
\begin{align}
\alpha+\beta+R_{s0} &\leq \min_{j=1,2}I(U;Y_j) \label{ach_gen_again_v11_first}\\
R_{s0} &\leq \min_{j=1,2}I(U;Y_j|Q)-I(U;Z|Q)\\
R_{s1}&\leq I(V_1;Y_1|U)-I(V_1;Z|U)\\
R_{s2}&\leq I(V_2;Y_2|U)-I(V_2;Z|U)\\
R_{s1}+R_{s2}&\leq
I(V_1;Y_1|U)+I(V_2;Y_2|U)-I(V_1;V_2|U)\nonumber\\
&\quad
-I(V_1,V_2;Z|U)\\
R_{p1}-\alpha+R_{s1}&\leq I(V_1;Y_1|U) \\
R_{p2}-\beta+R_{s2}&\leq I(V_2;Y_2|U)\\
R_{s1}+R_{p1}-\alpha+R_{s2}&\leq I(V_1;Y_1|U)+I(V_2;Y_2|U)-I(V_2;Z|U)\\
R_{s1}+R_{s2}+R_{p2}-\beta&\leq I(V_1;Y_1|U)+I(V_2;Y_2|U)-I(V_1;Z|U)\\
R_{s1}+R_{p1}-\alpha+R_{s2}+R_{p2}-\beta&\leq
I(V_1;Y_1|U)+I(V_2;Y_2|U)-I(V_1;V_2|U)\\
0&\leq \alpha\\
0&\leq \beta \\
\alpha&\leq R_{p1}\\
\beta &\leq R_{p2}
 \label{ach_gen_again_v11_last}
\end{align}
We eliminate $\alpha$ from the bounds in
(\ref{ach_gen_again_v11_first})-(\ref{ach_gen_again_v11_last}) by
using Fourier-Motzkin elimination, which leads to the following
region.
\begin{align}
\beta+R_{s0} &\leq \min_{j=1,2}I(U;Y_j) \label{ach_gen_again_v12_first}\\
R_{s0} &\leq \min_{j=1,2}I(U;Y_j|Q)-I(U;Z|Q)\\
R_{s1}&\leq I(V_1;Y_1|U)-I(V_1;Z|U)\\
R_{s2}&\leq I(V_2;Y_2|U)-I(V_2;Z|U)\\
R_{s1}+R_{s2}&\leq
I(V_1;Y_1|U)+I(V_2;Y_2|U)-I(V_1;V_2|U)\nonumber\\
&\quad
-I(V_1,V_2;Z|U)\\
R_{p2}-\beta+R_{s2}&\leq I(V_2;Y_2|U)\\
R_{s1}+R_{s2}+R_{p2}-\beta&\leq I(V_1;Y_1|U)+I(V_2;Y_2|U)-I(V_1;Z|U)\\
R_{s1}+R_{s2}+R_{p2}-\beta &\leq
I(V_1;Y_1|U)+I(V_2;Y_2|U)-I(V_1;V_2|U)\\
R_{s0}+\beta+R_{p1}+R_{s1}&\leq \min_{j=1,2}
I(U;Y_j)+I(V_1;Y_1|U)\\
\beta + R_{s0}+R_{s1}+R_{p1}+R_{s2}&\leq
\min_{j=1,2}I(U;Y_j)+I(V_1;Y_1|U)+I(V_2;Y_2|U)\nonumber\\
&\qquad\quad -I(V_2;Z|U)\\
R_{s0}+R_{s1}+R_{p1}+R_{s2}+R_{p2}&\leq
\min_{j=1,2}I(U;Y_j)+I(V_1;Y_1|U)+I(V_2;Y_2|U)\nonumber\\
&\qquad \quad -I(V_1;V_2|U)\\
0&\leq \beta \\
\beta &\leq R_{p2}
 \label{ach_gen_again_v12_last}
\end{align}
Next, we eliminate $\beta$ from the bounds in
(\ref{ach_gen_again_v12_first})-(\ref{ach_gen_again_v12_last}) by
using Fourier-Motzkin elimination, which leads to the following
region.
\begin{align}
R_{s0} &\leq \min_{j=1,2}I(U;Y_j|Q)-I(U;Z|Q)\label{ach_gen_again_v13_first}\\
R_{s1}&\leq I(V_1;Y_1|U)-I(V_1;Z|U)\\
R_{s2}&\leq I(V_2;Y_2|U)-I(V_2;Z|U)\\
R_{s1}+R_{s2}&\leq
I(V_1;Y_1|U)+I(V_2;Y_2|U)-I(V_1;V_2|U)\nonumber\\
&\quad
-I(V_1,V_2;Z|U)\\
R_{s0}+R_{s1}+R_{p1}&\leq \min_{j=1,2}I(U;Y_j)+I(V_1;Y_1|U)\\
R_{s0}+R_{s2}+R_{p2}&\leq \min_{j=1,2}I(U;Y_j)+I(V_2;Y_2|U)\\
R_{s0}+R_{s1}+R_{p1}+R_{s2}&\leq
\min_{j=1,2}I(U;Y_j)+I(V_1;Y_1|U)+I(V_2;Y_2|U) \nonumber\\
&\qquad\quad -I(V_2;Z|U)\\
R_{s0}+R_{s1}+R_{s2}+R_{p2}&\leq
\min_{j=1,2}I(U;Y_j)+I(V_1;Y_1|U)+I(V_2;Y_2|U)\nonumber\\
&\qquad \quad -I(V_1;Z|U)\\
R_{s0}+R_{s1}+R_{p1}+R_{s2}+R_{p2}&\leq
\min_{j=1,2}I(U;Y_j)+I(V_1;Y_1|U)+I(V_2;Y_2|U)\nonumber\\
&\qquad \quad -I(V_1;V_2|U)
 \label{ach_gen_again_v13_last}
\end{align}
We note the fact that since the common confidential message is
decoded by both legitimate users, its rate can be given up in
favor of each legitimate user's own confidential message rate
$R_{sj},~j=1,2,$ i.e., if the rate tuple
$(R_{p0},R_{s0},R_{p1},R_{s1},R_{p2},R_{s2})$ is achievable, the
rate tuple
$(R_{p0},R_{s0}-\alpha-\beta,R_{p1},R_{s1}+\alpha,R_{p2},R_{s2}+\beta)$
is also achievable for any non-negative $(\alpha,\beta)$ pairs
satisfying $\alpha+\beta\leq R_{s0}$. Using this fact for the
region given by
(\ref{ach_gen_again_v13_first})-(\ref{ach_gen_again_v13_last}), we
obtain the following region.
\begin{align}
\alpha+\beta &\leq \min_{j=1,2}I(U;Y_j|Q)-I(U;Z|Q)\label{ach_gen_again_v14_first}\\
R_{s1}-\alpha &\leq I(V_1;Y_1|U)-I(V_1;Z|U)\\
R_{s2}-\beta &\leq I(V_2;Y_2|U)-I(V_2;Z|U)\\
R_{s1}+R_{s2}-\alpha-\beta &\leq
I(V_1;Y_1|U)+I(V_2;Y_2|U)-I(V_1;V_2|U)
-I(V_1,V_2;Z|U)\\
\beta+R_{s1}+R_{p1}&\leq \min_{j=1,2}I(U;Y_j)+I(V_1;Y_1|U)\\
\alpha+R_{s2}+R_{p2}&\leq \min_{j=1,2}I(U;Y_j)+I(V_2;Y_2|U)\\
R_{s1}+R_{p1}+R_{s2}&\leq
\min_{j=1,2}I(U;Y_j)+I(V_1;Y_1|U)+I(V_2;Y_2|U)-I(V_2;Z|U)\\
R_{s1}+R_{s2}+R_{p2}&\leq
\min_{j=1,2}I(U;Y_j)+I(V_1;Y_1|U)+I(V_2;Y_2|U)-I(V_1;Z|U)\\
R_{s1}+R_{p1}+R_{s2}+R_{p2}&\leq
\min_{j=1,2}I(U;Y_j)+I(V_1;Y_1|U)+I(V_2;Y_2|U)-I(V_1;V_2|U)\\
0&\leq \alpha\\
0&\leq \beta \\
\alpha &\leq R_{s1}\\
\beta &\leq R_{s2}
 \label{ach_gen_again_v14_last}
\end{align}
We eliminate $\alpha$ from the bounds in
(\ref{ach_gen_again_v14_first})-(\ref{ach_gen_again_v14_last}) by
using Fourier-Motzkin elimination, which leads to the following
region.
\begin{align}
\beta &\leq \min_{j=1,2}I(U;Y_j|Q)-I(U;Z|Q)\label{ach_gen_again_v15_first}\\
R_{s1}+\beta &\leq
\min_{j=1,2}I(U;Y_j|Q)+I(V_1;Y_1|U)-I(U,V_1;Z|Q)\\
R_{s2}-\beta &\leq I(V_1;Y_1|U)+I(V_2;Y_2|U)-I(V_1;V_2|U)-I(V_1,V_2;Z|U)\\
R_{s2}-\beta &\leq I(V_2;Y_2|U)-I(V_2;Z|U)\\
R_{s1}+R_{s2}&\leq
\min_{j=1,2}I(U;Y_j|Q)+I(V_1;Y_1|U)+I(V_2;Y_2|U)-I(V_1;V_2|U)\nonumber\\
&\qquad\quad -I(U,V_1,V_2;Z|Q)\\
R_{s2}+R_{p2}&\leq \min_{j=1,2}I(U;Y_j)+I(V_2;Y_2|U)\\
\beta+R_{s1}+R_{p1}&\leq \min_{j=1,2}I(U;Y_j)+I(V_1;Y_1|U)\\
R_{s1}+R_{p1}+R_{s2}&\leq
\min_{j=1,2}I(U;Y_j)+I(V_1;Y_1|U)+I(V_2;Y_2|U)-I(V_2;Z|U)\\
R_{s1}+R_{s2}+R_{p2}&\leq
\min_{j=1,2}I(U;Y_j)+I(V_1;Y_1|U)+I(V_2;Y_2|U)-I(V_1;Z|U)\\
R_{s1}+R_{p1}+R_{s2}+R_{p2}&\leq
\min_{j=1,2}I(U;Y_j)+I(V_1;Y_1|U)+I(V_2;Y_2|U)-I(V_1;V_2|U)\\
R_{s1}+2R_{s2}+R_{p2}-\beta &\leq
\min_{j=1,2}I(U;Y_j)+I(V_1;Y_1|U)+2I(V_2;Y_2|U)-I(V_1;V_2|U)\nonumber\\
&\qquad \quad -I(V_1,V_2;Z|U)\\
0&\leq \beta \\
\beta &\leq R_{s2}\\
0&\leq I(V_1;Y_1|U)-I(V_1;Z|U)
 \label{ach_gen_again_v15_last}
\end{align}
We eliminate $\beta$ from the bounds in
(\ref{ach_gen_again_v15_first})-(\ref{ach_gen_again_v15_last}) by
using Fourier-Motzkin elimination, which leads to the following
region.
\begin{align}
R_{s1} &\leq
\min_{j=1,2}I(U;Y_j|Q)+I(V_1;Y_1|U)-I(U,V_1;Z|Q) \label{ach_gen_again_v16_first} \\
R_{s2}&\leq \min_{j=1,2}I(U;Y_j|Q)+I(V_2;Y_2|U)-I(U,V_2;Z|Q)\\
R_{s1}+R_{s2}&\leq
\min_{j=1,2}I(U;Y_j|Q)+I(V_1;Y_1|U)+I(V_2;Y_2|U)-I(V_1;V_2|U)\nonumber\\
&\qquad\quad -I(U,V_1,V_2;Z|Q)\\
R_{s1}+R_{p1}&\leq \min_{j=1,2}I(U;Y_j)+I(V_1;Y_1|U)\\
R_{s2}+R_{p2}&\leq \min_{j=1,2}I(U;Y_j)+I(V_2;Y_2|U)\\
R_{s1}+R_{p1}+R_{s2}&\leq
\min_{j=1,2}I(U;Y_j)+I(V_1;Y_1|U)+I(V_2;Y_2|U)-I(V_2;Z|U)\\
R_{s1}+R_{p1}+R_{s2}&\leq \min_{j=1,2}
I(U;Y_j)+2I(V_1;Y_1|U)+I(V_2;Y_2|U)-I(V_1;V_2|U) \nonumber\\
&\qquad \quad -I(V_1,V_2;Z|U)\\
R_{s1}+R_{s2}+R_{p2}&\leq
\min_{j=1,2}I(U;Y_j)+I(V_1;Y_1|U)+I(V_2;Y_2|U)-I(V_1;Z|U)\\
R_{s1}+R_{s2}+R_{p2}&\leq \min_{j=1,2}
I(U;Y_j)+I(V_1;Y_1|U)+2I(V_2;Y_2|U)-I(V_1;V_2|U) \nonumber\\
&\qquad \quad -I(V_1,V_2;Z|U)\\
R_{s1}+R_{p1}+R_{s2}+R_{p2}&\leq
\min_{j=1,2}I(U;Y_j)+I(V_1;Y_1|U)+I(V_2;Y_2|U)-I(V_1;V_2|U) \label{just_before_removal}\\
0&\leq \min_{j=1,2} I(U;Y_j|Q)-I(U;Z|Q)\label{remove_1}\\
0&\leq I(V_1;Y_1|U)-I(V_1;Z|U)\label{remove_2}\\
0&\leq I(V_2;Y_2|U)-I(V_2;Z|U)\label{remove_3}\\
0&\leq I(V_1;Y_1|U)+I(V_2;Y_2|U)-I(V_1;V_2|U)-I(V_1,V_2;Z|U)
 \label{ach_gen_again_v16_last}
\end{align}
Finally, we note that we can remove the bounds in
(\ref{remove_1})-(\ref{ach_gen_again_v16_last}) without enlarging
the region given by
(\ref{ach_gen_again_v16_first})-(\ref{just_before_removal}), which
will leave us with the desired achievable rate region given in
Theorem~\ref{theorem_inner_bound_general}; completing the proof.

\section{Proof of Lemma~\ref{lemma_eavesdropper_can_decode}}
\label{proof_eavesdropper_can_decode}

Assume that the eavesdropper tries to decode
$W_{p1},D_1,L_1,W_{p2},D_2,L_2$ by using its knowledge of
$(W_{s0},W_{s1},W_{s2},W_{p0})$. In particular, assume that, given
$(W_{s0}=w_{s0},W_{s1}=w_{s1},W_{s2}=w_{s1},W_{p0}=w_{p0})$, the
eavesdropper tries to decode $W_{p1},D_1,L_1,W_{p2},D_2,L_2$ by
looking for the unique $(V_1^n,V_2^n)$ such that
$(q^n,u^n,v_1^n,v_2^n,z^n)$ is jointly typical. There are four
possible error events:
\begin{itemize}
\item $\mathcal{E}^e_0=\{(q^n,u^n,v_1^n,v_2^n,z^n) $ is not
jointly typical for the transmitted $(q^n,u^n,v_1^n,v_2^n)\}$,

\item $\mathcal{E}^e_{i}=\{(W_{p1},D_1,L_1)\neq
(1,1,1),(W_{p2},D_2,L_2)\neq (1,1,1)$, and the corresponding tuple
$(q^n,u^n,v_1^n,v_2^n,z^n)$ is jointly typical$\}$,

\item $\mathcal{E}^e_{ii}=\{ (W_{p1},D_1,L_1)=
(1,1,1),(W_{p2},D_2,L_2)\neq (1,1,1)$, and the corresponding tuple
$(q^n,u^n,v_1^n,v_2^n,z^n)$ is jointly typical$\}$,

\item $\mathcal{E}^e_{iii}=\{(W_{p1},D_1,L_1)\neq
(1,1,1),(W_{p2},D_2,L_2)= (1,1,1)$, and the corresponding tuple
$(q^n,u^n,v_1^n,v_2^n,z^n)$ is jointly typical$\}$,

\end{itemize}
Thus, the probability of decoding error at the eavesdropper is
given by
\begin{align}
\Pr[\mathcal{E}^e]&=\Pr[\mathcal{E}^e_0\cup \mathcal{E}_{i}^e\cup
\mathcal{E}_{ii}^e
\cup \mathcal{E}_{iii}^e ] \\
&\leq\Pr[\mathcal{E}^e_0]+\Pr[\mathcal{E}_{i}^e]+\Pr[
\mathcal{E}_{ii}^e]+\Pr[ \mathcal{E}_{iii}^e]\\
&\leq \epsilon_{1n}+\Pr[\mathcal{E}_{i}^e]+\Pr[
\mathcal{E}_{ii}^e]+\Pr[ \mathcal{E}_{iii}^e]
\label{pe_at_eavesdropper}
\end{align}
where we first use the union bound, and next the fact that
$\Pr[\mathcal{E}^e_0]\leq \epsilon_{1n}$ for some $\epsilon_{1n}$
satisfying $\epsilon_{1n}\rightarrow 0$ as $n\rightarrow \infty$,
which follows from the properties of the jointly typical
sequences~\cite{cover_book}. Next, we consider the first term in
(\ref{pe_at_eavesdropper}) as follows
\begin{align}
\Pr[\mathcal{E}_{i}^e]&\leq \sum_{\substack{(w_{p1},d_1,l_1)\neq
(1,1,1)\\(w_{p2},d_2,l_2)\neq (1,1,1)}}
\Pr[(q^n,u^n,V_1^n,V_2^n,Z^n)\in\mathcal{A}_\epsilon^n]\\
&\leq \sum_{\substack{(w_{p1},d_1,l_1)\neq
(1,1,1)\\(w_{p2},d_2,l_2)\neq (1,1,1)}}
\sum_{(v_1^n,v_2^n,z^n)\in\mathcal{A}_\epsilon^n}
p(v_1^n|u^n)p(v_2^n|u^n)p(z^n|u^n)\label{generation_of_codewords_1}\\
&\leq \sum_{\substack{(w_{p1},d_1,l_1)\neq
(1,1,1)\\(w_{p2},d_2,l_2)\neq (1,1,1)}}
\sum_{(v_1^n,v_2^n,z^n)\in\mathcal{A}_\epsilon^n}
2^{-n(H(V_1|U)-\gamma_\epsilon)}2^{-n(H(V_2|U)-\gamma_\epsilon)}2^{-n(H(Z|U)-\gamma_\epsilon)}\label{typicality_implies_2}\\
&= \sum_{\substack{(w_{p1},d_1,l_1)\neq
(1,1,1)\\(w_{p2},d_2,l_2)\neq (1,1,1)}} |\mathcal{A}_\epsilon^n|
2^{-n(H(V_1|U)+H(V_2|U)+H(Z|U)-3\gamma_\epsilon)}\\
&\leq \sum_{\substack{(w_{p1},d_1,l_1)\neq
(1,1,1)\\(w_{p2},d_2,l_2)\neq (1,1,1)}}
2^{n(H(V_1,V_2,Z|U)+\gamma_\epsilon)}
2^{-n(H(V_1|U)+H(V_2|U)+H(Z|U)-3\gamma_\epsilon)}\label{typicality_implies_3}\\
&=\sum_{\substack{(w_{p1},d_1,l_1)\neq
(1,1,1)\\(w_{p2},d_2,l_2)\neq (1,1,1)}}
2^{-n(I(V_1,V_2;Z|U)+I(V_2;V_1|U)-4\gamma_\epsilon)}\\
&\leq 2^{n(R_{p1}+\Delta_1+L_1+R_{p2}+\Delta_2+L_2)}
2^{-n(I(V_1,V_2;Z|U)+I(V_2;V_1|U)-4\gamma_\epsilon)}
\label{pe_at_eavesdropper_2}
\end{align}

\noindent where $\mathcal{A}_{\epsilon}^n$ denotes the typical
set, $\gamma_\epsilon $ is a constant which is a function of
$\epsilon$, and satisfies $\gamma_\epsilon \rightarrow 0$ as
$\epsilon \rightarrow 0$, (\ref{generation_of_codewords_1}) comes
from the way we generated the sequences $(q^n,u^n,v_1^n,v_2^n)$,
(\ref{typicality_implies_2}) stems from the properties of the
typical sequences~\cite{cover_book}, and
(\ref{typicality_implies_3}) comes from the bounds on the size of
the typical set $\mathcal{A}_\epsilon^n$~\cite{cover_book}.
Equation (\ref{pe_at_eavesdropper_2}) implies that
$\Pr[\mathcal{E}_{i}^e]$ vanishes as $n\rightarrow \infty$ if the
following condition is satisfied.
\begin{align}
R_{p1}+\Delta_1+L_1+R_{p2}+\Delta_2+R_{p2}<
I(V_1,V_2;Z|U)+I(V_2;V_1|U)-4\gamma_\epsilon
\label{rate_constraint_for_eve_2}
\end{align}
Next, we consider $\Pr[ \mathcal{E}_{ii}^e]$ as follows
\begin{align}
\Pr[ \mathcal{E}_{ii}^e]&\leq\sum_{(w_{p2},d_2,l_2)\neq (1,1,1)}
\Pr[(q^n,u^n,v_1^n,V_2^n,Z^n)\in\mathcal{A}_{\epsilon}^n]\\
&\leq \sum_{(w_{p2},d_2,l_2)\neq (1,1,1)}
\sum_{(v_2^n,z^n)\in\mathcal{A}_{\epsilon}^n}p(v_2^n|u^n)p(z^n|u^n,v_1^n)\label{generation_of_codewords_2}\\
&\leq  \sum_{(w_{p2},d_2,l_2)\neq (1,1,1)}
\sum_{(v_2^n,z^n)\in\mathcal{A}_{\epsilon}^n}
2^{-n(H(V_2|U)-\gamma_\epsilon)}
2^{-n(H(Z|U,V_1)-\gamma_\epsilon)}\label{typicality_implies_4}\\
&=\sum_{(w_{p2},d_2,l_2)\neq (1,1,1)} |\mathcal{A}_{\epsilon}^n|
2^{-n(H(V_2|U)-\gamma_\epsilon)}
2^{-n(H(Z|U,V_1)-\gamma_\epsilon)} \\
&\leq \sum_{(w_{p2},d_2,l_2)\neq (1,1,1)}
2^{n(H(V_2,Z|U,V_1)+\gamma_\epsilon)}
2^{-n(H(V_2|U)-\gamma_\epsilon)}
2^{-n(H(Z|U,V_1)-\gamma_\epsilon)} \label{typicality_implies_5}\\
&= \sum_{(w_{p2},d_2,l_2)\neq (1,1,1)}
2^{-n(I(V_2;Z,V_1|U)-3\gamma_\epsilon)}\\
&\leq 2^{n(R_{p2}+\Delta_2+L_2)}
2^{-n(I(V_2;Z,V_1|U)-3\gamma_\epsilon)}
\label{pe_at_eavesdropper_3}
\end{align}
where (\ref{generation_of_codewords_2}) comes from the way we
generated sequences $(q^n,u^n,v_1^n,v_2^n)$,
(\ref{typicality_implies_4}) is due to the properties of the
typical sequences~\cite{cover_book}, and
(\ref{typicality_implies_5}) comes from the bounds on the typical
set $\mathcal{A}_\epsilon^n$~\cite{cover_book}. Equation
(\ref{pe_at_eavesdropper_3}) implies that $\Pr[
\mathcal{E}_{ii}^e] \rightarrow 0$ as $n\rightarrow \infty$ if the
following condition is satisfied.
\begin{align}
R_{p2}+\Delta_2+L_2<I(V_2;Z,V_1|U)-3\gamma_\epsilon
\label{rate_constraint_for_eve_3}
\end{align}
Similarly, we can show that $\Pr[ \mathcal{E}_{iii}^e] \rightarrow
0$ as $n\rightarrow \infty$ if the following condition is
satisfied.
\begin{align}
R_{p1}+\Delta_1+L_1<I(V_1;Z,V_2|U)-3\gamma_\epsilon
\label{rate_constraint_for_eve_4}
\end{align}
Thus, we have shown that if the rates
$(R_{p1},\Delta_1,L_1,R_{p2},\Delta_2,L_2)$ satisfy
(\ref{rate_constraint_for_eve_2}),
(\ref{rate_constraint_for_eve_3}),
(\ref{rate_constraint_for_eve_4}), the eavesdropper can decode
$W_{p1},D_1,L_1,W_{p2},D_2,L_2$ by using its knowledge of
$(W_{s0},W_{s1},W_{s2},\break W_{p0})$, i.e., $\Pr[\mathcal{E}^e]$
vanishes as $n\rightarrow \infty$. In view of this fact, using
Fano's lemma, we get
\begin{align}
H(W_{p1},D_1,L_1,W_{p2},D_2,L_2|Z^n,Q^n,W_{s0},W_{s1},W_{s2},W_{p0},D_0)\leq
n\gamma_{2n}
\end{align}
where $\gamma_{2n}\rightarrow 0$ as $n\rightarrow \infty$;
completing the proof.

\section{Proofs of Theorems~\ref{theorem_suff_1}~and~\ref{theorem_outer_Gauss}}
\label{proofs_of_degraded_Gaussian_MIMO}

\subsection{Background}
\label{sec:background} We need some properties of the Fisher
information and the differential entropy, which are provided next.

\begin{Def}[\!\!\cite{MIMO_BC_Secrecy}, Definition~3]
Let $(\bbu,\bbx)$ be an arbitrarily correlated length-$n$ random
vector pair with well-defined densities. The conditional Fisher
information matrix of $\bbx$ given $\bbu$ is defined as
\begin{align}
\bbj(\bbx|\bbu)=E\left[\brho(\bbx|\bbu)\brho(\bbx|\bbu)^\top\right]
\end{align}
where the expectation is over the joint density $f(\bu,\bx)$, and
the conditional score function $\brho(\bx|\bu)$ is
\begin{align}
\brho(\bx|\bu)&=\nabla \log f(\bx|\bu)=\left[~\frac{\partial\log
f(\bx|\bu)}{\partial x_1}~~\ldots~~\frac{\partial\log
f(\bx|\bu)}{\partial x_n}~\right]^\top
\end{align}
\end{Def}

We first present the conditional form of the Cramer-Rao
inequality, which is proved in~\cite{MIMO_BC_Secrecy}.
\begin{Lem}[\!\!\cite{MIMO_BC_Secrecy}, Lemma~13]
\label{lemma_conditional_crb_vector} Let $\bbu,\bbx$ be
arbitrarily correlated random vectors with well-defined densities.
Let the conditional covariance matrix of $\bbx$ be ${\rm
Cov}(\bbx|\bbu)\succ \bzero$, then we have
\begin{align}
\bbj(\bbx|\bbu)\succeq {\rm Cov}(\bbx|\bbu)^{-1}
\end{align}
which is satisfied with equality if $(\bbu,\bbx)$ is jointly
Gaussian with conditional covariance matrix ${\rm
Cov}(\bbx|\bbu)$.
\end{Lem}

The following lemma will be used in the upcoming proof. The
unconditional version of this lemma, i.e., the case $\bbt=\phi$,
is proved in Lemma 6 of~\cite{MIMO_BC_Secrecy}.
\begin{Lem}[\!\!\cite{MIMO_BC_Secrecy},~Lemma~6]
\label{lemma_change_in_fisher} Let $\bbt,\bbu,\bbv_1,\bbv_2$ be
random vectors such that $(\bbt,\bbu)$ and \break$(\bbv_1,\bbv_2)$
are independent. Moreover, let $\bbv_1,\bbv_2$ be Gaussian random
vectors with covariance matrices $\bbsigma_1,\bbsigma_2$ such that
$\bzero \prec \bbsigma_1 \preceq \bbsigma_2$. Then, we have
\begin{align}
\bbj^{-1}(\bbu+\bbv_2|\bbt)-\bbsigma_2 \succeq
\bbj^{-1}(\bbu+\bbv_1|\bbt)-\bbsigma_1
\end{align}
\end{Lem}

The following lemma addresses the change of the Fisher information with respect to conditioning.
\begin{Lem} [\!\!\cite{MIMO_BC_Secrecy}, Lemma~17]
\label{lemma_cond_increases_fi} Let $(\bbv,\bbu,\bbx)$ be
length-$n$ random vectors with well-defined densities. Moreover,
assume that the partial derivatives of $f(\bu|\bx,\bv)$ with
respect to $x_i,~i=1,\ldots,n$ exist and satisfy
\begin{align}
\max_{1\leq i\leq n}\left|\frac{\partial f(\bu|\bv,\bx)}{\partial
x_i}\right| \leq g(\bu)
\end{align}
for some integrable function $g(\bu)$. Then, if $(\bbu,\bbv,\bbx)$
satisfy the Markov chain $\bbu\rightarrow \bbv \rightarrow \bbx$,
we have
\begin{align}
\bbj(\bbx|\bbv) \succeq \bbj(\bbx|\bbu)
\end{align}
\end{Lem}

The following lemma will also be used in the upcoming proof.
\begin{Lem}[\!\!\cite{MIMO_BC_Secrecy},~Lemma~8]
\label{Shamai_s_lemma} Let $\bbk_1,\bbk_2$ be positive
semi-definite matrices satisfying
$\bzero\preceq\bbk_1\preceq\bbk_2$, and $\mathbf{f}(\bbk)$ be a
matrix-valued function such that $\mathbf{f}(\bbk)\succeq\bzero$
for $\bbk_1\preceq\bbk\preceq \bbk_2$. Moreover, $\mathbf{f}(\bbk)$ is assumed to be gradient
of some scalar field. Then, we have
\begin{align}
\int_{\bbk_1}^{\bbk_2}\mathbf{f}(\bbk)d\bbk \geq 0
\end{align}
\end{Lem}

The following generalization of the de Bruijn identity~\cite{Stam,
Blachman} is due to~\cite{Palomar_Gradient}, where the
unconditional form of this identity, i.e., $\bbu=\phi$, is proved.
Its generalization to this conditional form for an arbitrary
$\bbu$ is rather straightforward, and is given in Lemma~16
of~\cite{MIMO_BC_Secrecy}.
\begin{Lem}[\!\!\cite{MIMO_BC_Secrecy}, Lemma~16]
\label{gradient_fisher_conditional} Let $(\bbu,\bbx)$ be an
arbitrarily correlated random vector pair with finite second order
moments, and also be independent of the random vector $\bbn$ which
is zero-mean Gaussian with covariance matrix
$\bbsigma_N\succ\bzero$. Then, we have
\begin{align}
\nabla_{\bbsigma_N} h(\bbx+\bbn|\bbu)=\frac{1}{2}
\bbj(\bbx+\bbn|\bbu)
\end{align}
\end{Lem}

The following lemma is due to~\cite{Dembo,Dembo_Cover} which lower
bounds the differential entropy in terms of the Fisher information
matrix.
\begin{Lem}[\!\!\cite{Dembo,Dembo_Cover}]
\label{lemma_dembo} Let $(U,\bbx)$ be an $(n+1)$-dimensional
random vector, where the conditional Fisher information matrix of
$\bbx$, conditioned on $U$, exists. Then, we have
\begin{align}
h(\bbx|U) \geq \frac{1}{2} \log |(2\pi e)\bbj^{-1}(\bbx|U)|
\end{align}
\end{Lem}

We also need the following fact about strictly positive definite matrices.
\begin{Lem}
\label{lemma_order_change}
Let $\bba,\bbb$ be two positive definite matrices, i.e., $\bba \succ \bzero,\bbb \succ \bzero$. If $\bba \preceq
\bbb$, we have $\bba^{-1}\succeq \bbb^{-1}$.
\end{Lem}

\subsection{Proofs}

First, we prove Theorem~\ref{theorem_suff_1} by showing that for
any $(U,V,\bbx)$, there exists a Gaussian $(U^{G},V^{G},\bbx^{G})$
which provides a larger region. Essentially, this proof will also
yield a proof for Theorem~\ref{theorem_outer_Gauss} because the
outer bound in Theorem~\ref{theorem_outer_bound} is defined by the
same inequalities that define the inner bound given in
Theorem~\ref{theorem_inner_bound} except the inequality in
(\ref{extra_bound}). Thus, we only provide the proof of
Theorem~\ref{theorem_suff_1}.

\vspace{0.25cm}

\noindent \underline{First step:} We consider the bound on
$R_{s2}$ given in (\ref{final_bounds_first}) as follows
\begin{align}
R_{s2} &\leq I(U;\bby_2)-I(U;\bbz)\\
&=[h(\bby_2)-h(\bbz)]+[h(\bbz|U)-h(\bby_2|U)] \label{bound_rs2}
\end{align}
First, we consider the first term in (\ref{bound_rs2}) as follows
\begin{align}
h(\bbz)-h(\bby_2)=\frac{1}{2} \int_{\bbsigma_2}^{\bbsigma_Z}
\bbj(\bbx+\bbn) d\bbsigma_N \label{bound_rs2_part1}
\end{align}
which follows from Lemma~\ref{gradient_fisher_conditional}, and
$\bbn$ is a Gaussian random vector with covariance matrix
$\bbsigma_N$ satisfying $\bbsigma_2\preceq \bbsigma_N \preceq
\bbsigma_Z$. We note that due to
Lemma~\ref{lemma_conditional_crb_vector}, we have
\begin{align}
\bbj(\bbx+\bbn)\succeq ({\rm Cov}(\bbx)+\bbsigma_N)^{-1}\succeq
(\bbs+\bbsigma_N)^{-1}\label{order_k2_prime_implies}
\end{align}
where the second inequality comes from
Lemma~\ref{lemma_order_change} and the fact that
$E\left[\bbx\bbx^\top\right]\preceq \bbs$. Using this inequality
in (\ref{bound_rs2_part1}), we get
\begin{align}
h(\bbz)-h(\bby_2)&\geq \frac{1}{2} \int_{\bbsigma_2}^{\bbsigma_Z}
(\bbs+\bbsigma_N)^{-1} d\bbsigma_N
\label{lemma_integral_implies} \\
&=\frac{1}{2} \log \frac{|\bbs+\bbsigma_Z|}{|\bbs+\bbsigma_2|}
\label{bound_rs2_part1_final}
\end{align}
where (\ref{lemma_integral_implies}) comes from Lemma~\ref{Shamai_s_lemma}.

Next, we consider the second term in (\ref{bound_rs2}) as follows
\begin{align}
h(\bbz|U)-h(\bby_2|U)=\frac{1}{2} \int_{\bbsigma_2}^{\bbsigma_Z}
\bbj(\bbx+\bbn|U) d\bbsigma_N \label{bound_rs2_part2}
\end{align}
which follows from Lemma~\ref{gradient_fisher_conditional}, and $\bbn$ is a Gaussian random vector
with covariance matrix $\bbsigma_N$ satisfying $\bbsigma_2\preceq \bbsigma_N \preceq \bbsigma_Z$. Using
Lemma~\ref{lemma_change_in_fisher}, for any $\bbsigma_N$ satisfying $\bbsigma_2\preceq \bbsigma_N\preceq \bbsigma_Z$,
we have
\begin{align}
\bbj^{-1}(\bbx+\bbn_2|U)-\bbsigma_2 \preceq
\bbj^{-1}(\bbx+\bbn|U)-\bbsigma_N \preceq
\bbj^{-1}(\bbx+\bbn_Z|U)-\bbsigma_Z
\label{lemma_change_in_fisher_implies}
\end{align}
Due to Lemma~\ref{lemma_order_change}, the inequalities in (\ref{lemma_change_in_fisher_implies})
imply
\begin{align}
\left[\bbj^{-1}(\bbx+\bbn_Z|U)-\bbsigma_Z+\bbsigma_N \right]^{-1}
\preceq \bbj(\bbx+\bbn|U) \preceq
\left[\bbj^{-1}(\bbx+\bbn_2|U)-\bbsigma_2+\bbsigma_N\right]^{-1}
\end{align}
Using these inequalities in (\ref{bound_rs2_part2}) in conjunction with Lemma~\ref{Shamai_s_lemma},
we get
\begin{align}
\frac{1}{2} \log
\frac{|\bbj^{-1}(\bbx+\bbn_Z|U)|}{|\bbj^{-1}(\bbx+\bbn_Z)-\bbsigma_Z+\bbsigma_2|}
\leq h(\bbz|U)-h(\bby_2|U) \leq \frac{1}{2} \log
\frac{|\bbj^{-1}(\bbx+\bbn_2|U)-\bbsigma_2+\bbsigma_Z|}{|\bbj^{-1}(\bbx+\bbn_2)|}
\end{align}
which can be expressed as
\begin{align}
f(0)&\leq h(\bbz|U)-h(\bby_2|U) \leq f(1)
\end{align}
where $f(t)$ is defined as
\begin{align}
f(t)=\frac{1}{2} \log \frac{|\bbk_1(t)+\bbsigma_Z|}{|\bbk_1(t)+\bbsigma_2|},\quad 0\leq t\leq 1
\end{align}
and $\bbk_1(t)$ is given by
\begin{align}
\bbk_1
(t)=(1-t)\left[\bbj^{-1}(\bbx+\bbn_Z|U)-\bbsigma_Z\right]+t\left[\bbj^{-1}(\bbx+\bbn_2|U)-\bbsigma_2\right]
\label{k1_t}
\end{align}
Since $f(t)$ is continuous in $t$, due to the intermediate value theorem, there exists a $t_1^*$
such that $0\leq t_1^* \leq 1$, and
\begin{align}
f(t_1^*)&=h(\bbz|U)-h(\bby_2|U)\\
&=\frac{1}{2} \log\frac{|\bbk_1+\bbsigma_Z|}{|\bbk_1+\bbsigma_2|}
\label{bound_rs2_part2_final}
\end{align}
where $\bbk_1=\bbk_1(t_1^*)$. Since $0\leq t_1^* \leq 1$, $\bbk_1$ satisfies
\begin{align}
\bbj^{-1}(\bbx+\bbn_2|U)-\bbsigma_2 \preceq \bbk_1 \preceq
\bbj^{-1}(\bbx+\bbn_Z|U)-\bbsigma_Z \label{order_k1_first}
\end{align}
in view of (\ref{k1_t}). Moreover, we have
\begin{align}
\bbk_1 &\preceq \bbj^{-1}(\bbx+\bbn_Z|U)-\bbsigma_Z \label{conditioning_0}\\
&\preceq \bbj^{-1}(\bbx+\bbn_Z)-\bbsigma_Z \label{conditioning_1} \\
&\preceq {\rm Cov}(\bbx+\bbn_Z)-\bbsigma_Z \label{crb}\\
&\preceq {\rm Cov}(\bbx)\\
&\preceq \bbs \label{order_k2_prime_implies_1}
\end{align}
where (\ref{conditioning_1}) is due to
Lemma~\ref{lemma_cond_increases_fi} and (\ref{crb}) comes from
Lemma~\ref{lemma_conditional_crb_vector}. Thus, in view of
(\ref{order_k1_first}) and (\ref{order_k2_prime_implies_1}),
$\bbk_1$ satisfies
\begin{align}
\bbj^{-1}(\bbx+\bbn_2|U)-\bbsigma_2 \preceq \bbk_1 \preceq \bbs
\label{order_k1}
\end{align}
Now, using (\ref{bound_rs2_part1_final}) and (\ref{bound_rs2_part2_final}) in (\ref{bound_rs2}),
we get the following bound on $R_{s2}$
\begin{align}
R_{s2} \leq \frac{1}{2} \log
\frac{|\bbs+\bbsigma_2|}{|\bbk_1+\bbsigma_2|} -\frac{1}{2} \log
\frac{|\bbs+\bbsigma_Z|}{|\bbk_1+\bbsigma_Z|}
\end{align}
which completes the first step of the proof.

\vspace{0.25cm}

\noindent \underline{Second step:} We consider the bound on the
confidential message sum rate $R_{s1}+R_{s2}$ given in
(\ref{final_bounds_3}) as follows
\begin{align}
R_{s1}+R_{s2} & \leq I(U;\bby_2)+I(\bbx;\bby_1|U)-I(\bbx;\bbz) \\
&=\left[h(\bby_2)-h(\bbz)\right]+\left[h(\bby_1|U)-h(\bby_2|U)\right]-\frac{1}{2} \log \frac{|\bbsigma_1|}{|\bbsigma_Z|} \\
&\leq \frac{1}{2} \log \frac{|\bbs+\bbsigma_2|}{|\bbs+\bbsigma_Z|}
+\left[h(\bby_1|U)-h(\bby_2|U)\right]-\frac{1}{2} \log
\frac{|\bbsigma_1|}{|\bbsigma_Z|}
\label{bound_rs2_part1_final_implies}
\end{align}
where (\ref{bound_rs2_part1_final_implies}) comes from (\ref{bound_rs2_part1_final}). Next, we consider the remaining term
in (\ref{bound_rs2_part1_final_implies}) as follows
\begin{align}
h(\bby_2|U)-h(\bby_1|U)&=\frac{1}{2}
\int_{\bbsigma_1}^{\bbsigma_2} \bbj(\bbx+\bbn|U) d\bbsigma_N
\label{gradient_fisher_conditional_implies}
\end{align}
which follows from Lemma~\ref{gradient_fisher_conditional}, and $\bbn$ is a Gaussian random vector with covariance
matrix $\bbsigma_N$ satisfying $\bbsigma_1 \preceq \bbsigma_N \preceq \bbsigma_2$. Due to Lemma~\ref{lemma_change_in_fisher},
for any Gaussian random vector $\bbn$ with $\bbsigma_N\preceq \bbsigma_2$,
we have
\begin{align}
\bbj^{-1}(\bbx+\bbn|U)-\bbsigma_N &\preceq \bbj^{-1}(\bbx+\bbn_2|U)-\bbsigma_2 \\
&\preceq \bbk_1 \label{order_k1_implies}
\end{align}
where (\ref{order_k1_implies}) comes from (\ref{order_k1}). In view of Lemma~\ref{lemma_order_change},
(\ref{order_k1_implies}) implies
\begin{align}
\bbj(\bbx+\bbn|U) \succeq (\bbk_1+\bbsigma_N)^{-1},\quad
\bbsigma_N \preceq \bbsigma_2 \label{order_k1_implies_1}
\end{align}
Using (\ref{order_k1_implies_1}) in (\ref{gradient_fisher_conditional_implies})
in conjunction with Lemma~\ref{Shamai_s_lemma}, we have
\begin{align}
h(\bby_2|U)-h(\bby_1|U) &\geq \frac{1}{2} \int_{\bbsigma_1}^{\bbsigma_2} (\bbk_1+\bbsigma_N)^{-1} d\bbsigma_N \\
&=\frac{1}{2} \log \frac{|\bbk_1+\bbsigma_2|}{|\bbk_1+\bbsigma_1|} \label{bound_rs1_rs2_remainder}
\end{align}
Using (\ref{bound_rs1_rs2_remainder}) in (\ref{bound_rs2_part1_final_implies}), we get
\begin{align}
R_{s1}+R_{s2} \leq \frac{1}{2} \log
\frac{|\bbs+\bbsigma_2|}{|\bbk_1+\bbsigma_2|} +\frac{1}{2} \log
\frac{|\bbk_1+\bbsigma_1|}{|\bbsigma_1|} -\frac{1}{2} \log
\frac{|\bbs+\bbsigma_Z|}{|\bbsigma_Z|}
\end{align}
which completes the second step of the proof.

\vspace{0.25cm} \noindent \underline{Third step:} We consider the
bound on $R_{s2}+R_{p2}$ given in (\ref{final_bounds_4}) as
follows
\begin{align}
R_{p2}+R_{s2}&\leq I(U;\bby_2) \\
&=h(\bby_2)-h(\bby_2|U)\\
&\leq \frac{1}{2} \log |(2\pi e)(\bbs+\bbsigma_2)|-h(\bby_2|U)
\label{bound_rs2_part1_final_implies_1}
\end{align}
where (\ref{bound_rs2_part1_final_implies_1}) comes from the
maximum entropy theorem~\cite{cover_book}. Next, we consider the
remaining term in (\ref{bound_rs2_part1_final_implies_1}). Using
(\ref{bound_rs2_part2_final}), we have
\begin{align}
h(\bby_2|U)&=h(\bbz|U) -\frac{1}{2} \log \frac{|\bbk_1+\bbsigma_Z|}{|\bbk_1+\bbsigma_2|}\\
&\geq \frac{1}{2} \log |(2\pi
e)\bbj^{-1}(\bbx+\bbn_Z|U)|-\frac{1}{2} \log
\frac{|\bbk_1+\bbsigma_Z|}{|\bbk_1+\bbsigma_2|}
\label{lemma_dembo_implies}\\
&\geq \frac{1}{2} \log |(2\pi e) (\bbk_1+\bbsigma_Z)|-\frac{1}{2} \log \frac{|\bbk_1+\bbsigma_Z|}{|\bbk_1+\bbsigma_2|}
\label{order_k1_implies_2} \\
&=\frac{1}{2} \log |(2\pi e) (\bbk_1+\bbsigma_2)| \label{lower_hy2u}
\end{align}
where (\ref{lemma_dembo_implies}) is due to Lemma~\ref{lemma_dembo},
and (\ref{order_k1_implies_2}) comes from (\ref{conditioning_0}) and monotonicity of
$|\cdot|$ in positive semi-definite matrices. Using (\ref{lower_hy2u})
in (\ref{bound_rs2_part1_final_implies_1}), we get
\begin{align}
R_{p2}+R_{s2} &\leq \frac{1}{2} \log
\frac{|\bbs+\bbsigma_2|}{|\bbk_1+\bbsigma_2|}
\end{align}
which completes the third step of the proof.

\vspace{0.25cm} \noindent \underline{Fourth step:} We consider the
bound in (\ref{extra_bound}) as follows
\begin{align}
\lefteqn{R_{s1}+R_{p2}+R_{s2} \leq I(U;\bby_2)+I(\bbx;\bby_1|U)-I(\bbx;\bbz|U)} \\
&=h(\bby_2)-h(\bby_2|U)+\left[h(\bby_1|U)-h(\bbz|U)\right]-\frac{1}{2} \log\frac{|\bbsigma_1|}{|\bbsigma_Z|}\\
&\leq \frac{1}{2} \log |(2\pi e) (\bbs+\bbsigma_2)|-h(\bby_2|U)+\left[h(\bby_1|U)-h(\bbz|U)\right] -\frac{1}{2} \log\frac{|\bbsigma_1|}{|\bbsigma_Z|} \label{max_entropy_implies_2} \\
&\leq \frac{1}{2} \log |(2\pi e) (\bbs+\bbsigma_Z)|
-\frac{1}{2}\log |(2\pi e)(\bbk_1+\bbsigma_2)|+\left[h(\bby_1|U)-h(\bbz|U)\right]\nonumber\\
&\quad -\frac{1}{2} \log\frac{|\bbsigma_1|}{|\bbsigma_Z|} \label{lower_hy2u_implies}\\
& = \frac{1}{2} \log \frac{|\bbs+\bbsigma_2|}{
|\bbk_1+\bbsigma_2|}+\left[h(\bby_1|U)-h(\bby_2|U)\right]+\left[h(\bby_2|U)-h(\bbz|U)\right]
 -\frac{1}{2} \log\frac{|\bbsigma_1|}{|\bbsigma_Z|} \\
&= \frac{1}{2} \log \frac{|\bbs+\bbsigma_2|}{
|\bbk_1+\bbsigma_2|}+\left[h(\bby_1|U)-h(\bby_2|U)\right]
+\frac{1}{2} \log\frac{|\bbk_1+\bbsigma_2|}{|\bbk_1+\bbsigma_Z|}
-\frac{1}{2} \log\frac{|\bbsigma_1|}{|\bbsigma_Z|}
\label{bound_rs2_part1_final_implies_3}\\
& \leq \frac{1}{2} \log\frac{ |\bbs+\bbsigma_2|}{
|\bbk_1+\bbsigma_2|}+\frac{1}{2}\log\frac{|\bbk_1+\bbsigma_1|}{|\bbk_1+\bbsigma_2|}+\frac{1}{2}
\log\frac{|\bbk_1+\bbsigma_2|}{|\bbk_1+\bbsigma_Z|} -\frac{1}{2}
\log\frac{|\bbsigma_1|}{|\bbsigma_Z|}
\label{bound_rs1_rs2_remainder_implies}\\
&=\frac{1}{2} \log \frac{|\bbs+\bbsigma_2|}{|\bbk_1+\bbsigma_2|}
+\frac{1}{2} \log
\frac{|\bbk_1+\bbsigma_1|}{|\bbsigma_1|}-\frac{1}{2} \log
\frac{|\bbk_1+\bbsigma_Z|}{|\bbsigma_Z|}
\end{align}
where (\ref{max_entropy_implies_2}) comes from the maximum entropy
theorem~\cite{cover_book}, (\ref{lower_hy2u_implies}) comes from
(\ref{lower_hy2u}), (\ref{bound_rs2_part1_final_implies_3}) is due
to (\ref{bound_rs2_part2_final}), and
(\ref{bound_rs1_rs2_remainder_implies}) comes from
(\ref{bound_rs1_rs2_remainder}).

\vspace{0.25cm} \noindent \underline{Fifth step:} We consider the
bound in (\ref{final_bounds_last}) as follows
\begin{align}
\lefteqn{R_{p1}+R_{s1}+R_{p2}+R_{s2} \leq I(U;\bby_2)+I(\bbx;\bby_1|U)} \\
&=h(\bby_2)+\left[h(\bby_1|U)-h(\bby_2|U)\right]-\frac{1}{2} \log |(2\pi e)\bbsigma_1| \\
&\leq \frac{1}{2} \log |(2\pi
e)(\bbs+\bbsigma_2)|+\left[h(\bby_1|U)-h(\bby_2|U)\right]-\frac{1}{2}
\log |(2\pi e)\bbsigma_1|
\label{max_entropy_implies_3}\\
&\leq \frac{1}{2} \log |(2\pi e)(\bbs+\bbsigma_2)|
+\frac{1}{2}\log \frac{|\bbk_1+\bbsigma_1|}{|\bbk_1+\bbsigma_2|}-\frac{1}{2} \log |(2\pi e)\bbsigma_1| \label{bound_rs1_rs2_remainder_implies_1}\\
&=\frac{1}{2} \log \frac{|\bbs+\bbsigma_2|}{|\bbk_1+\bbsigma_2|}
+\frac{1}{2} \log \frac{|\bbk_1+\bbsigma_1|}{|\bbsigma_1|}
\end{align}
where (\ref{max_entropy_implies_3}) comes from the maximum entropy
theorem~\cite{cover_book}, and
(\ref{bound_rs1_rs2_remainder_implies_1}) comes from
(\ref{bound_rs1_rs2_remainder}).

Hence, we have shown that for any feasible $(U,\bbx)$, there
exists a Gaussian $(U^{G},\bbx^G)$ which yields a larger rate
region. This completes the proof.

\section{Proof of Theorem~\ref{theorem_ach_general_Gaussian_MIMO}}
\label{proof_of_theorem_ach_general_Gaussian_MIMO}

We now obtain an alternative rate region by using the one given by
(\ref{ach_gen_again_v16_first})-(\ref{ach_gen_again_v16_last}).
This alternative region is more amenable for evaluation for the
Gaussian MIMO channel. We note that the following region is
included in the region given by
(\ref{ach_gen_again_v16_first})-(\ref{ach_gen_again_v16_last}).
\begin{align}
R_{s1} &\leq \min_{j=1,2}I(U;Y_j|Q)+I(V_1;Y_1|U)-I(V_1;V_2|U)-I(U;Z|Q)\nonumber\\
&\qquad \quad -I(V_1;Z|U,V_2) \label{ach_gen_MIMO_again_v1_first}\\
R_{s2} &\leq \min_{j=1,2} I(U;Y_j|Q)+ I(V_2;Y_2|U)-I(U,V_2;Z|Q) \\
R_{s1}+R_{s2}&\leq \min_{j=1,2}
I(U;Y_j|Q)+I(V_1;Y_1|U)+I(V_2;Y_2|U)-I(V_1;V_2|U)\nonumber\\
&\qquad \quad -I(U,V_1,V_2;Z|Q)\\
R_{s1}+R_{p1} &\leq \min_{j=1,2} I(U;Y_j)+I(V_1;Y_1|U)-I(V_1;V_2|U)\\
R_{s2}+R_{p2}&\leq \min_{j=1,2}I(U;Y_j)+I(V_2;Y_2|U)\\
R_{s1}+R_{p1}+R_{s2} &\leq \min_{j=1,2} I(U;Y_j)+I(V_1;Y_1|U)+I(V_2;Y_2|U)-I(V_1;V_2|U)\nonumber\\
&\qquad \quad -I(V_2;Z|U)\\
R_{s1}+R_{s2}+R_{p2}&\leq \min_{j=1,2}I(U;Y_j)+I(V_1;Y_1|U)+I(V_2;Y_2|U)-I(V_1;V_2|U)\nonumber\\
&\qquad \quad -I(V_1;Z|U,V_2)\\
R_{s1}+R_{p1}+R_{s2}+R_{p2}&\leq
\min_{j=1,2}I(U;Y_j)+I(V_1;Y_1|U)+I(V_2;Y_2|U)-I(V_1;V_2|U)\label{just_before_removal_mimo_1}\\
0&\leq \min_{j=1,2} I(U;Y_j|Q)-I(U;Z|Q)\label{remove_mimo_1}\\
0&\leq I(V_1;Y_1|U)-I(V_1;V_2|U)-I(V_1;Z|U,V_2)\\
0&\leq I(V_2;Y_2|U)-I(V_2;Z|U) \label{ach_gen_MIMO_again_v1_last}
\end{align}
We note that we can remove the constraints given by
(\ref{remove_mimo_1})-(\ref{ach_gen_MIMO_again_v1_last}) without
enlarging the region given by
(\ref{ach_gen_MIMO_again_v1_first})-(\ref{just_before_removal_mimo_1}),
which will leave us with the following region.
\begin{align}
R_{s1} &\leq \min_{j=1,2}I(U;Y_j|Q)+I(V_1;Y_1|U)-I(V_1;V_2|U)-I(U;Z|Q)\nonumber\\
&\qquad\quad -I(V_1;Z|U,V_2) \label{ach_gen_MIMO_again_v2_first}\\
R_{s2} &\leq \min_{j=1,2} I(U;Y_j|Q)+ I(V_2;Y_2|U)-I(U,V_2;Z|Q) \\
R_{s1}+R_{s2}&\leq \min_{j=1,2}
I(U;Y_j|Q)+I(V_1;Y_1|U)+I(V_2;Y_2|U)-I(V_1;V_2|U)\nonumber\\
&\qquad \quad -I(U,V_1,V_2;Z|Q)\\
R_{s1}+R_{p1} &\leq \min_{j=1,2} I(U;Y_j)+I(V_1;Y_1|U)-I(V_1;V_2|U)\\
R_{s2}+R_{p2}&\leq \min_{j=1,2}I(U;Y_j)+I(V_2;Y_2|U)\\
R_{s1}+R_{p1}+R_{s2} &\leq \min_{j=1,2} I(U;Y_j)+I(V_1;Y_1|U)+I(V_2;Y_2|U)-I(V_1;V_2|U)\nonumber\\
&\qquad \quad -I(V_2;Z|U)\\
R_{s1}+R_{s2}+R_{p2}&\leq \min_{j=1,2}I(U;Y_j)+I(V_1;Y_1|U)+I(V_2;Y_2|U)-I(V_1;V_2|U)\nonumber\\
&\qquad \quad -I(V_1;Z|U,V_2)\\
R_{s1}+R_{p1}+R_{s2}+R_{p2}&\leq
\min_{j=1,2}I(U;Y_j)+I(V_1;Y_1|U)+I(V_2;Y_2|U)-I(V_1;V_2|U)
\label{ach_gen_MIMO_again_v2_last}
\end{align}
We denote the region given by
(\ref{ach_gen_MIMO_again_v2_first})-(\ref{ach_gen_MIMO_again_v2_last})
by $\mathcal{R}_{21}$. Similarly, the following achievable rate
region can be obtained as well.
\begin{align}
R_{s1} &\leq \min_{j=1,2} I(U;Y_j|Q)+ I(V_1;Y_1|U)-I(U,V_1;Z|Q)\label{ach_gen_MIMO_again_v3_first} \\
R_{s2} &\leq \min_{j=1,2}I(U;Y_j|Q)+I(V_2;Y_2|U)-I(V_1;V_2|U)-I(U;Z|Q)\nonumber\\
&\qquad \quad -I(V_2;Z|U,V_1) \\
R_{s1}+R_{s2}&\leq \min_{j=1,2}
I(U;Y_j|Q)+I(V_1;Y_1|U)+I(V_2;Y_2|U)-I(V_1;V_2|U)\nonumber\\
&\qquad \quad -I(U,V_1,V_2;Z|Q)\\
R_{s1}+R_{p1} &\leq \min_{j=1,2} I(U;Y_j)+I(V_1;Y_1|U)\\
R_{s2}+R_{p2}&\leq \min_{j=1,2}I(U;Y_j)+I(V_2;Y_2|U)-I(V_1;V_2|U)\\
R_{s1}+R_{p1}+R_{s2} &\leq \min_{j=1,2} I(U;Y_j)+I(V_1;Y_1|U)+I(V_2;Y_2|U)-I(V_1;V_2|U)\nonumber\\
&\qquad \quad -I(V_2;Z|U,V_1)\\
R_{s1}+R_{s2}+R_{p2}&\leq \min_{j=1,2}I(U;Y_j)+I(V_1;Y_1|U)+I(V_2;Y_2|U)-I(V_1;V_2|U)\nonumber\\
&\qquad \quad -I(V_1;Z|U)\\
R_{s1}+R_{p1}+R_{s2}+R_{p2}&\leq
\min_{j=1,2}I(U;Y_j)+I(V_1;Y_1|U)+I(V_2;Y_2|U)-I(V_1;V_2|U)
\label{ach_gen_MIMO_again_v3_last}
\end{align}
which is denoted by $\mathcal{R}_{12}$. Hence, we obtain the
achievable rate region $\mathcal{R}$ which is given by
\begin{align}
\mathcal{R}={\rm conv} \left(\mathcal{R}_{12}\cup
\mathcal{R}_{21}\right)
\end{align}
Next, we outline an alternative method to obtain the region
$\mathcal{R}_{21}$. First, we set $L_1,L_2,R_{p1},\break
R_{p2},\Delta_1,\Delta_2$ as follows.
\begin{align}
L_1&=I(V_1;V_2|U)\label{selection_for_MIMO_1}\\
L_2&=0 \label{selection_for_MIMO_2} \\
R_{p1}+\Delta_1&=I(V_1;Z|U,V_2)\label{selection_for_MIMO_3} \\
R_{p2}+\Delta_2&= I(V_2;Z|U) \label{selection_for_MIMO_4}
\end{align}
Using the values of $L_1,L_2,R_{p1}+\Delta_1,R_{p2}+\Delta_2$
given by (\ref{selection_for_MIMO_1})-(\ref{selection_for_MIMO_4})
in (\ref{ach_gen_again_v1_first})-(\ref{ach_gen_again_v1_last}),
we have the following achievable rate region.
\begin{align}
R_{p0}+R_{s0}+\Delta_0 &\leq \min_{j=1,2} I(U;Y_j)\label{ach_gen_MIMO_alt_v1_first}\\
\tilde{\tilde{R}}_{p0}+R_{s0}+\Delta_0 &\leq
\min_{j=1,2}I(U;Y_j|Q)\\
R_{s1}&\leq I(V_1;Y_1|U)-I(V_1;V_2|U)-I(V_1;Z|U,V_2)\\
R_{s2}&\leq I(V_2;Y_2|U)-I(V_2;Z|U)\\
\tilde{\tilde{R}}_{p0}+\Delta_0&=I(U;Z|Q)\\
R_{p1}+\Delta_1&=I(V_1;Z|U,V_2)\\
R_{p2}+\Delta_2&=I(V_2;Z|U) \label{ach_gen_MIMO_alt_v1_last}
\end{align}
Next, following the procedure in
Appendix~\ref{sec:Fourier-Motzkin}, the achievable rate region
given by
(\ref{ach_gen_MIMO_again_v2_first})-(\ref{ach_gen_MIMO_again_v2_last}),
i.e., $\mathcal{R}_{21}$, can be obtained by using the achievable
rate region given by
(\ref{ach_gen_MIMO_alt_v1_first})-(\ref{ach_gen_MIMO_alt_v1_last}).
Similarly, the other region $\mathcal{R}_{12}$ can be obtained as
well. We also note that this alternative derivation reveals that
since we select $L_1=I(V_1;V_2|U),L_2=0$ to obtain the achievable
rate region $\mathcal{R}_{21}$, in this case, the transmitter
first encodes $V_2^n$, and then, next using the non-causal
knowledge of $V_2^n$, encodes $V_1^n$, i.e., uses Gelfand-Pinsker
encoding for $V_2^n$.

Next, we obtain an achievable rate region for the Gaussian MIMO
multi-receiver wiretap channel with public and confidential
messages. We provide this achievable rate region by evaluating the
regions $\mathcal{R}_{12}$ and $\mathcal{R}_{21}$ with a specific
choice of $Q,U,V_1,V_2,\bbx$. In particular, to evaluate
$\mathcal{R}_{21}$, we use the following selection for
$Q,U,V_1,V_2,\bbx$:
\begin{itemize}
\item $Q$ is selected as a zero-mean Gaussian random vector with
covariance matrix $\bbs-\bbk_0-\bbk_1-\bbk_2$, where
$\bbk_0,\bbk_1,\bbk_2$ are positive semi-definite matrices
satisfying $\bbk_0+\bbk_1+\bbk_2 \preceq \bbs$,

\item $U$ is selected as $U=Q+Q^\prime$, where $Q^\prime$ is a
zero-mean Gaussian random vector with covariance matrix $\bbk_0$,
and is independent of $Q$,

\item $V_2$ is selected as $V_2=U+U_2$, where $U_2$ is a zero-mean
Gaussian random vector with covariance matrix $\bbk_2$, and is
independent of $Q,Q^\prime$,

\item $V_1$ is selected as $V_1=U_1+\bba U_2+U$, where $U_1$ is a
zero-mean Gaussian random vector with covariance matrix $\bbk_1$,
and is independent of $Q,Q^\prime,U_2$. The encoding matrix $\bba$
is given by $\bba=\bbk_1\left[\bbk_1+\bbsigma_1\right]^{-1}$,

\item $\bbx$ is selected as $\bbx=Q+Q^\prime+U_2+U_1$.
\end{itemize}

We note that we use dirty-paper coding~\cite{Wei_Yu} to encode
$V_1$, which leads to the following.
\begin{align}
I(V_1;\bby_1|U)-I(V_1;V_2|U)=\frac{1}{2} \log
\frac{|\bbk_1+\bbsigma_1|}{|\bbsigma_1|}
\end{align}
The other mutual information terms in the region
$\mathcal{R}_{21}$ can be computed straightforwardly, which leads
to the following region.
\begin{align}
R_{s1}&\leq \min_{j=1,2}\frac{1}{2} \log
\frac{|\bbk_0+\bbk_1+\bbk_2+\bbsigma_j|}{|\bbk_1+\bbk_2+\bbsigma_j|}
+\frac{1}{2}\log
\frac{|\bbk_1+\bbsigma_1|}{|\bbsigma_1|}\nonumber\\
&\qquad \quad- \frac{1}{2} \log
\frac{|\bbk_0+\bbk_1+\bbk_2+\bbsigma_Z|}{|\bbk_1+\bbk_2+\bbsigma_Z|}
-\frac{1}{2} \log \frac{|\bbk_1+\bbsigma_Z|}{|\bbsigma_Z|}\label{ach_general_Gauss_proof_first}\\
R_{s2}&\leq \min_{j=1,2} \frac{1}{2} \log
\frac{|\bbk_0+\bbk_1+\bbk_2+\bbsigma_j|}{|\bbk_1+\bbk_2+\bbsigma_j|}+
\frac{1}{2} \log
\frac{|\bbk_1+\bbk_2+\bbsigma_2|}{|\bbk_1+\bbsigma_2|}\nonumber\\
&\qquad \quad -\frac{1}{2}\log
\frac{|\bbk_0+\bbk_1+\bbk_2+\bbsigma_Z|}{|\bbk_1+\bbsigma_Z|}\\
R_{s1}+R_{s2}&\leq \min_{j=1,2}\frac{1}{2} \log
\frac{|\bbk_0+\bbk_1+\bbk_2+\bbsigma_j|}{|\bbk_1+\bbk_2+\bbsigma_j|}
+\frac{1}{2} \log
\frac{|\bbk_1+\bbk_2+\bbsigma_2|}{|\bbk_1+\bbsigma_2|}\nonumber\\
&\qquad \quad  +\frac{1}{2}\log
\frac{|\bbk_1+\bbsigma_1|}{|\bbsigma_1|} -\frac{1}{2}\log
\frac{|\bbk_0+\bbk_1+\bbk_2+\bbsigma_Z|}{|\bbsigma_Z|}\\
R_{s1}+R_{p1}&\leq \min_{j=1,2} \frac{1}{2} \log
\frac{|\bbs+\bbsigma_j|}{|\bbk_1+\bbk_2+\bbsigma_j|}+\frac{1}{2}\log
\frac{|\bbk_1+\bbsigma_1|}{|\bbsigma_1|}\\
R_{s2}+R_{p2}&\leq \min_{j=1,2} \frac{1}{2} \log
\frac{|\bbs+\bbsigma_j|}{|\bbk_1+\bbk_2+\bbsigma_j|}+\frac{1}{2}\log
\frac{|\bbk_1+\bbk_2+\bbsigma_2|}{|\bbk_1+\bbsigma_2|}\\
R_{s1}+R_{p1}+R_{s2}&\leq \min_{j=1,2} \frac{1}{2} \log
\frac{|\bbs+\bbsigma_j|}{|\bbk_1+\bbk_2+\bbsigma_j|}+\frac{1}{2}\log
\frac{|\bbk_1+\bbsigma_1|}{|\bbsigma_1|}\nonumber\\
&\qquad \quad +\frac{1}{2}\log
\frac{|\bbk_1+\bbk_2+\bbsigma_2|}{|\bbk_1+\bbsigma_2|}
-\frac{1}{2}\log
\frac{|\bbk_1+\bbk_2+\bbsigma_Z|}{|\bbk_1+\bbsigma_Z|}\\
R_{s1}+R_{s2}+R_{p2}&\leq \min_{j=1,2} \frac{1}{2} \log
\frac{|\bbs+\bbsigma_j|}{|\bbk_1+\bbk_2+\bbsigma_j|}+\frac{1}{2}\log
\frac{|\bbk_1+\bbk_2+\bbsigma_2|}{|\bbk_1+\bbsigma_2|}\nonumber\\
&\qquad \quad +\frac{1}{2}\log
\frac{|\bbk_1+\bbsigma_1|}{|\bbsigma_1|} -\frac{1}{2}\log
\frac{|\bbk_1+\bbsigma_Z|}{|\bbsigma_Z|}\\
R_{s1}+R_{p1}+R_{s2}+R_{p2}&\leq \min_{j=1,2} \frac{1}{2}\log
\frac{|\bbs+\bbsigma_j|}{|\bbk_1+\bbk_2+\bbsigma_j|}+\frac{1}{2}\log
\frac{|\bbk_1+\bbk_2+\bbsigma_2|}{|\bbk_1+\bbsigma_2|}\nonumber\\
&\qquad \quad +\frac{1}{2}\log
\frac{|\bbk_1+\bbsigma_1|}{|\bbsigma_1|}\label{ach_general_Gauss_proof_last}
\end{align}
We denote the region given in
(\ref{ach_general_Gauss_proof_first})-(\ref{ach_general_Gauss_proof_last})
by $\mathcal{R}_{21}(\bbk_0,\bbk_1,\bbk_2)$. Similarly, we can
evaluate the region $\mathcal{R}_{12}$ to obtain another
achievable rate region $\mathcal{R}_{12}(\bbk_0,\bbk_1,\bbk_2)$
for the Gaussian MIMO multi-receiver wiretap channel, where
$\mathcal{R}_{12}(\bbk_0,\bbk_1,\bbk_2)$ can be obtained from
$\mathcal{R}_{21}(\bbk_0,\bbk_1,\bbk_2)$ by swapping the
subscripts 1 and 2.

\bibliographystyle{unsrt}
\bibliography{IEEEabrv,references2}
\end{document}